\documentclass[sigconf]{acmart}
\AtBeginDocument{%
  }

\setcopyright{acmlicensed}
\copyrightyear{2018}
\acmYear{2018}
\acmDOI{XXXXXXX.XXXXXXX}
\acmConference[Conference acronym 'XX]{Make sure to enter the correct
  conference title from your rights confirmation email}{June 03--05,
  2018}{Woodstock, NY}
\acmISBN{978-1-4503-XXXX-X/2018/06}

\usepackage{booktabs}
\usepackage{multirow}
\usepackage{enumitem}
\usepackage{titlesec}
\usepackage{caption}
\usepackage{subcaption}
\usepackage{float}
\usepackage{makecell}
\usepackage{algorithm} 
\usepackage{algpseudocode} 

\newcommand{\ysrev}[2]{\textcolor{red}{#2}}

\begin{document}

\title{Request-Only Optimization for Recommendation Systems}


\author{
Liang Guo, 
Wei Li, 
Lucy Liao,
Huihui Cheng,
Rui Zhang,
Yu Shi,
Yueming Wang,
Yanzun Huang,
Keke Zhai,
Pengchao Wang,
Timothy Shi,
Xuan Cao,
Shengzhi Wang,
Renqin Cai,
Zhaojie Gong,
Omkar Vichare,
Rui Jian,
Leon Gao$^{\dagger}$,
Shiyan Deng,
Xingyu Liu,
Xiong Zhang,
Fu Li$^{\dagger}$,
Wenlei Xie$^{\dagger}$,
Bin Wen$^{\dagger}$,
Rui Li,
Lu Fang,
Xing Liu,
Jiaqi Zhai$^{\dagger}$
} 
\affiliation{Meta Platforms, Inc.
\country{USA}}

\renewcommand{\thefootnote}{\fnsymbol{footnote}}

\renewcommand{\shortauthors}{Guo et al.}
\renewcommand{\authors}{Liang Guo, Wei Li, Lucy Liao, Huihui Cheng, Rui Zhang, Yu Shi, Yueming Wang, Yanzun Huang, Keke Zhai, Pengchao Wang, Timothy Shi, Xuan Cao, Shengzhi Wang, Renqin Cai, Zhaojie Gong, Omkar Vichare, Rui Jian, Leon Gao, Shiyan Deng, Xingyu Liu, Xiong Zhang, Fu Li, Wenlei Xie, Bin Wen, Rui Li, Lu Fang, Xing Liu, Jiaqi Zhai}

\begin{abstract}

Recommendation systems represent one of the largest machine learning applications on the planet -- industry-scale recommendation models are trained with petabytes of data and serve billions of users every day. To utilize the rich user signals in the long user history, these models have been scaled up to unprecedented complexity, up to trillions of floating-point operations (TFLOPs) per example. This scale, coupled with the huge amount of training data, necessitates new storage and training algorithms to efficiently improve the quality of these complex recommendation systems. 

In this paper, we present a Request-Only Optimizations (ROO) training and modeling paradigm. ROO simultaneously improves the storage and training efficiency as well as the model quality of recommendation systems. We holistically approach this challenge through co-designing data (i.e., request-only data), infrastructure (i.e., request-only based data processing pipeline), and model architectures. Our ROO training and modeling paradigm treats a user request as a unit of the training data. Compared with the established practice of treating a user impression as a unit, this new design directly removes redundant features in data logging, saving data storage. Second, by de-duplicating computations and communications across multiple impressions in a request, this new paradigm enables highly scaled-up neural network architectures to better capture user interest signals, such as Generative Recommenders (GRs) and other request-only friendly architectures.

Our proposed ROO training and modeling paradigm has been deployed to three major recommendation products, each with billions of active users. ROO data format allows for increasing the training sample volume by 43\% to 150\% across these three products. ROO training optimization yields a substantial increase in training throughput -- up to 570\% for the retrieval and early-stage ranking models, and between 32\% and 100\% for the late-stage ranking models. Combined with ROO-based neural architectures like Hierarchical Sequential Transduction Units (HSTU), ROO scales model FLOPs by 7x utilizing the same amount of training compute, enabling significant offline and online metric wins. Our studies offer practical and scalable design solutions for engineers seeking to build efficient and effective large-scale recommendation systems.

\end{abstract}



\received{20 February 2007}
\received[revised]{12 March 2009}
\received[accepted]{5 June 2009}

\maketitle
\makeatletter
\gdef\@shortauthors{Guo et al.} 
\gdef\@authors{Liang Guo, Wei Li, Lucy Liao, Huihui Cheng, Rui Zhang, Yu Shi, Yueming Wang, Yanzun Huang, Keke Zhai, Pengchao Wang, Timothy Shi, Xuan Cao, Shengzhi Wang, Renqin Cai, Zhaojie Gong, Omkar Vichare, Rui Jian, Leon Gao, Shiyan Deng, Xingyu Liu, Xiong Zhang, Fu Li, Wenlei Xie, Bin Wen, Rui Li, Lu Fang, Xing Liu, Jiaqi Zhai} 
\makeatother
\footnotetext[2]{$^{\dagger}$Work done at Meta.}

\section{Introduction}
Deep Learning Recommendation Models (DLRMs) \cite{naumov2019deep,ytdnn_goog_recsys16} are the backbone of recommendation systems serving billions of users daily. The scale at which these models operate creates substantial computational and infrastructure challenges across the entire pipeline -- from data generation and pre-processing to GPU training and inference~\cite{jouppi2017datacenter,gupta2020architectural,sigcomm24_rdma_meta}. These challenges necessitate new solutions that reduce costs while maintaining or improving model quality.

\begin{figure*} 
  \centering
  \includegraphics[width=0.9\textwidth]{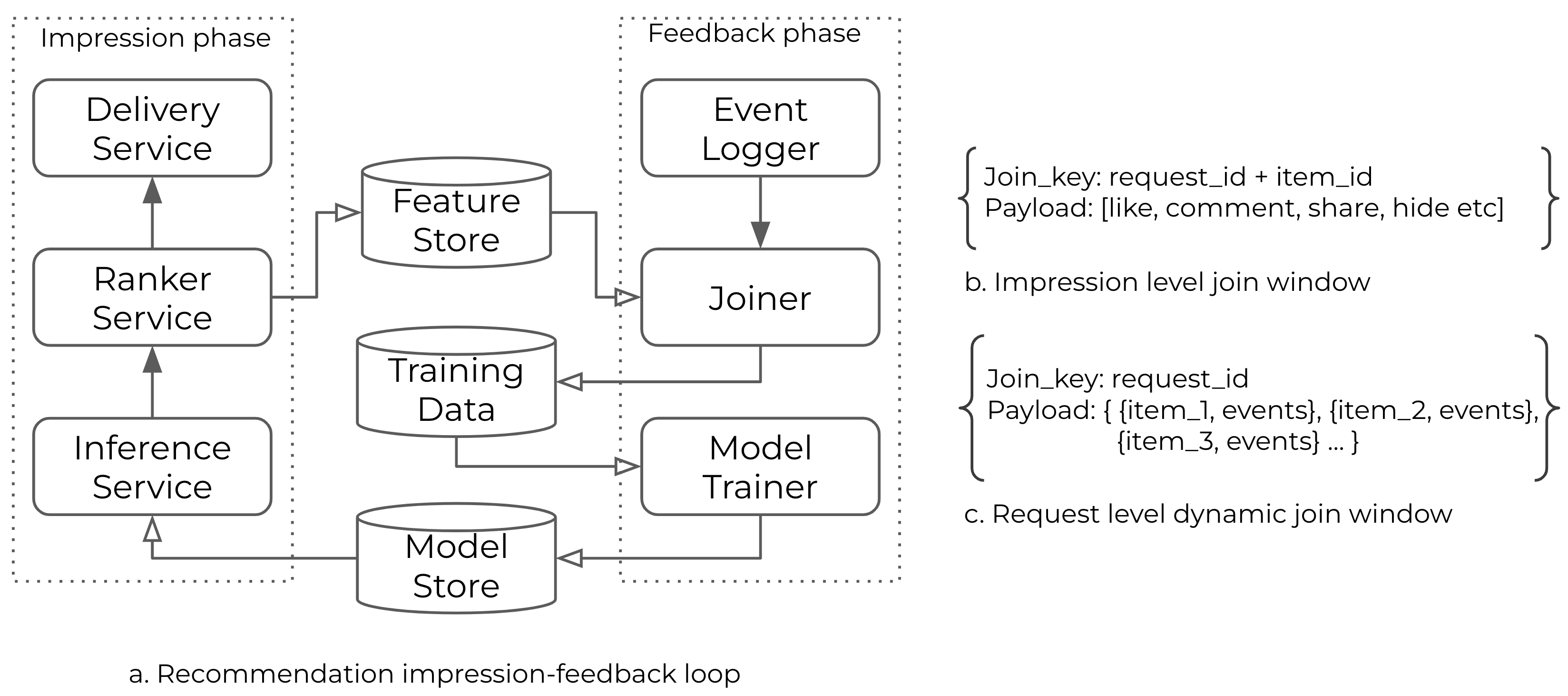}
  \caption{A traditional training data pipeline incurs excessive feature duplication at the source of data generation, i.e. impression-level event-feature joiner (b), wasting significant computational resources throughout the system. A request-level joiner (this work) replaces the traditional joiner and captures all impressions belonging to the same request in one single sample (c).} 
  \Description{}
  \label{fig:reco-datapipeline}
\end{figure*}

\begin{figure}
    \centering
    \includegraphics[width=0.9\linewidth]{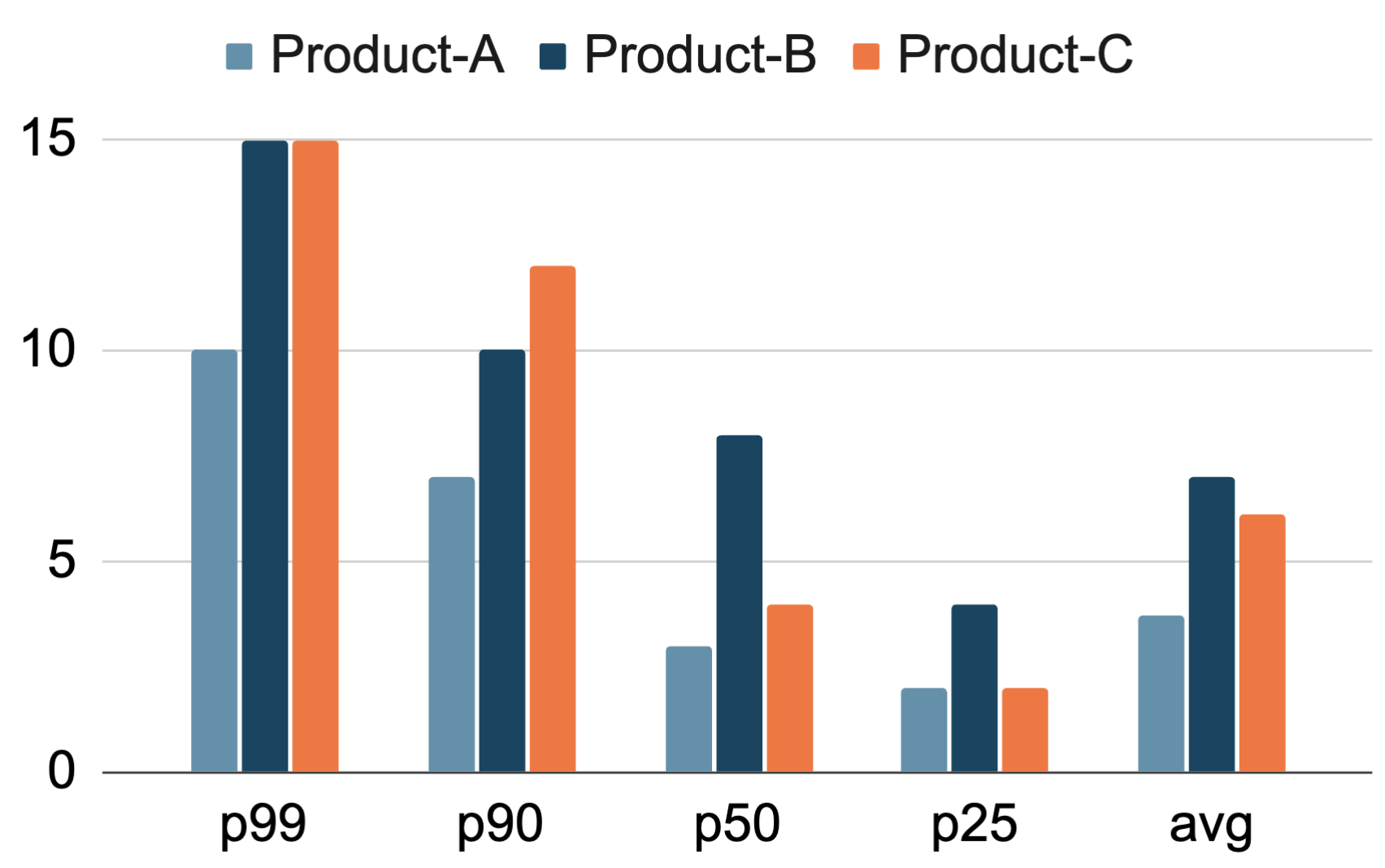}
    \caption{Number of items (i.e., number of impression-level samples) within requests across three major recommendation products on a billion-user-scale platform. }
    \vspace{-.3em}
    \label{fig:impression_per_req}
    \vspace{-.8em}
\end{figure}

A fundamental inefficiency in current recommendation systems lies in their data collection and processing pipelines. As illustrated in Figure~\ref{fig:reco-datapipeline}, traditional systems follow a two-phase approach: an impression phase where multiple content items are sent in response to a user request, followed by a feedback phase where user actions are collected and joined with features to create training samples.

The critical issue emerges at the impression-level event-feature joiner (Figure~\ref{fig:reco-datapipeline}.b), where extensive feature duplication occurs. For each user request that generates multiple impressions, identical user features are redundantly stored across all impression-level training samples. Figure~\ref{fig:impression_per_req} reveals that industrial recommendation products typically generate 4-7 impression samples per request. This redundancy cascades throughout the system -- consuming storage, network bandwidth, and GPU computation resources. 

This inefficiency is particularly pronounced for modern recommendation models that leverage long user history sequences. These sparse user features constitute the majority of features in training samples, as demonstrated in previous work~\cite{naumov2019deep}. The growing adoption of sequential modeling techniques further exacerbates this challenge, as these approaches rely heavily on the very user sequence features being redundantly processed~\cite{zhai2024actions,liu2024kuaiformer,pancha2022pinnerformer,kang2018self}. 

While request-level optimizations has been recently applied to model inference~\cite{you2024exploiting} to exploit the  single-user, multiple-candidates characteristic of recommendation model inference, extending this concept to training presents significant challenges. Existing implementations often act as wrappers around impression-level data at inference time, increasing system complexity. Potential training-time optimization is complicated by the widespread use of impression-based data formats throughout client and server infrastructures, and the incremental arrival of impression-based events during the feedback phase.

Our work presents a novel training data format, Request-Only Optimization (ROO), that captures all impression-level samples corresponding to the same request in a single training sample. This data format distinctly separates request-only (RO) data from non-request-only, or impression-level (NRO) data. This clean separation enables user sequence tensors to be processed exactly once across all impressions from the same request or app session, eliminating redundant computation of user-side features.

Building on this foundation, we develop ROO training methodologies and ROO-friendly model architectures that capitalize on this data structure. By processing user-side information at the request level rather than the impression level, we reduce the computational workload from $B_{NRO}$ to $B_{RO}$ examples (where $B_{NRO}$ represents the number of impression items in a batch and $B_{RO}$ represents the number of requests). With an average of 4-7 (Figure~\ref{fig:impression_per_req}) impressions per request ($B_{NRO} / B_{RO}$) and user-side features dominating the feature space, our approach yields theoretical speedups of 300-600\%.

Our ROO training paradigm delivers substantial benefits for retrieval and early-stage ranking models, which utilize minimal cross-features. These models demonstrate up to 6x improvements in training throughput with minimal code changes. For late-stage ranking models, we leverage ROO amortization to apply user feature compression, resulting in significant quality gains without increased computational costs.
Most importantly, the dramatic reduction in GPU training costs has enabled the adoption of previously prohibitive modeling technologies such as Generative Recommenders (GRs)~\cite{zhai2024actions}. This represents a fundamental shift in what becomes computationally feasible in production recommendation systems.

Overall, our paper makes the following key contributions:
\begin{itemize}
\item A new Request-Only Optimization (ROO) training data format that eliminates pervasive feature duplication at the source (Section~\ref{sec:roo-design}), increasing training sample volumes by 43\% to 150\% across different recommendation products utilizing the same storage capacity.

\item A new ROO training paradigm that maximizes recommendation system efficiency through computation redundancy elimination and at the same time, unifying the data format used across training and inference tech stacks.

\item Novel neural model architectures including UserArch, ROO sequential models, and HSTU that fully leverage the cost amortization benefits of ROO training (Section~\ref{sec:roo-model-archs}).

\item Extensive evaluation of ROO-based data formats and architectures throughout feature preprocessing, training, and inference across multiple ranking/retrieval stages for billion-user scale products, with experimental validation demonstrating significant efficiency gains and quality improvements in production (Section~\ref{sec:experiments}). 
\end{itemize}

\section{End-to-End ROO Design}
\label{sec:roo-design}
In this work, we take a first-principled approach and address the fundamental problem at the root level of data for DLRM training and inference. The end-to-end Request-Only Optimization training paradigm benefits from a cohesive co-design with data, system, and model architecture. The data format eliminates the feature duplication at the source and enables us to hatch a systematical implementation throughout storage, GPU resource usage, training, and inference. The synergy between ROO and advanced modeling techniques, such as HSTU~\cite{zhai2024actions}, strikes a great balance between model complexity and efficiency.

\subsection{Request-Level Training Data}
\label{sec:roo:request-level-training-data}
We redesign the training sample format as request level data format. In the recommendation feedback phase, a request level joiner is used to buffer user-item interaction events as grouped by unique request ids (Figure~\ref{fig:reco-datapipeline}.c). When the join window closes, it creates a compact ROO training example that has one copy of user features in the RO part and an array of item features in the NRO part (Table~\ref{tab:request_level_sample}). Besides the request level joiner, there is no additional infrastructure change in the training data pipeline. This is because the design eliminates user feature redundancy at the root data level and furthermore it fosters efficiency optimizations in the upstream model feature preprocessing and training pipeline.

\begin{table*}[t]
    \centering
    \begin{tabular}{|c|c|c|c|c|} \hline 
         request-id&  conversions&  dense features&  id-list features& id-score-list features\\ \hline 
 \verb|int|& \verb|list<int>|& \verb|map<int, float>|& \verb|map<int, float>| &\verb|map<int, map<int, float>>|\\ \hline
    \end{tabular}
    \vspace{.2em}
    \caption{Impression-level training sample schema}
    \label{tab:impression_sample}
\end{table*}

\begin{table*}[t]
    \centering
    \begin{tabular}{|c|c|c|c|c|c|} \hline 
         request-id&  conversions&  user dense features&  user id-list features&  item dense features& item id-list features\\ \hline 
         \verb|int| &  \verb|list<list<int>>|&  \verb|map<int,float>|&  \verb|map<int,list<int>>|&  \verb|map<int,list<float>>|& \verb|map<int,list<list<int>>|\\ \hline
    \end{tabular}
    \vspace{.2em}
    \caption{Request level training sample schema}
    \label{tab:request_level_sample}
\end{table*}

\subsubsection{Request-Level Data Ingestion}
The ROO training data schema (Table \ref{tab:request_level_sample}) allows RO and NRO features to be organized separately in feature flattening storage format \cite{zhao2022understanding} that enables better columnar storage compression ratio. ROO training samples are typically read in a mini-batch by data preprocessing (DPP) workers. Features are read from feature flattened tables into columnar major in-memory format for further feature preprocessing and tensor transformation. In each mini-batch, RO features are transformed into a 2D tensor representing RO user float features, along with RO user ID-list features structured as a PyTorch KeyedJaggedTensor \cite{ivchenko2022torchrec}. NRO features are similarly processed, with flattening along the candidate dimension to form an 2D NRO item float feature tensor and an NRO item ID-list KeyedJaggedTensor. The batch sizes for RO and NRO features, denoted as $B_{RO}$ and $B_{NRO}$ respectively, are inherently distinct. To avoid redundant duplication of user features for batch size alignment, a NRO tensor is used to specify the number of impressions per ROO training sample, thereby minimizing data copying and transformation of duplicated user features. Furthermore, tensor operations such as dense normalization, sequence merging, deduplication, and masking are optimized to enhance computational efficiency and performance. The efficiency of the ROO data schema extends from storage to DPP workers and greatly scales training data ingestion for DLRM and GR model training needs.

\subsubsection{ROO Data Activity-to-Serving Latency}
In recommendation products, activity-to-serving (ATS) latency is used to measure personalization models’ freshness, and activity-to-training latency is a critical part of it. In theory, adopting request-level joining  could  potentially increase activity-to-training latency because it needs to wait for multiple items within the request join window. In production, request-level joiners use either fixed-time policy or dynamic training trigger to close the join window before the fixed-time window closes. Through measurements, we establish that the average time gap between the earliest user-item interaction event arriving at the request join window buffer and the last item of the request is within 16 minutes. In practice, ROO training samples' data landing latency is about the half the fixed-time join window.

\subsubsection{Request Level Join Data Quality}
Changing training data joining from impression-level to request-level could theoretically lead to training data distribution changes which might have an impact on the recommendation model quality. However, our data quality analysis shows user training sample and feature coverage parity with impression-level training data. Table~\ref{tab:quality} shows that distributions of important labels such as conversion and video view duration have negligible mismatch.

\begin{table}[]
    \centering
    \begin{tabular}{cc}
        \toprule
         Label & Mismatch Rate  \\
         \midrule
        Conversion Mismatch Product-A & 0.03\% \\
        Conversion Mismatch Product-B & 0.16\% \\
        Conversion Mismatch Product-C & 0.023\% \\
        Video View Duration Mismatch Product-A & 0.035\% \\
        Video View Duration Mismatch Product-B & 1.07\% \\
        Video View Duration Mismatch Product-C & 0.01\% \\
        \bottomrule
    \end{tabular}
    \vspace{.3em}
    \caption{Request level join data quality as compared to impression level data}
    \label{tab:quality}
\end{table}

\subsection{Unifying Model Training and Inference Request-Level Optimization}
The ROO feature preprocessing ensures the same model input format for both training and inference stacks. This opens up engineering opportunities to simplify the DLRM system with significant performance improvements.

\begin{figure} 
  \centering
  \includegraphics[width=\linewidth]{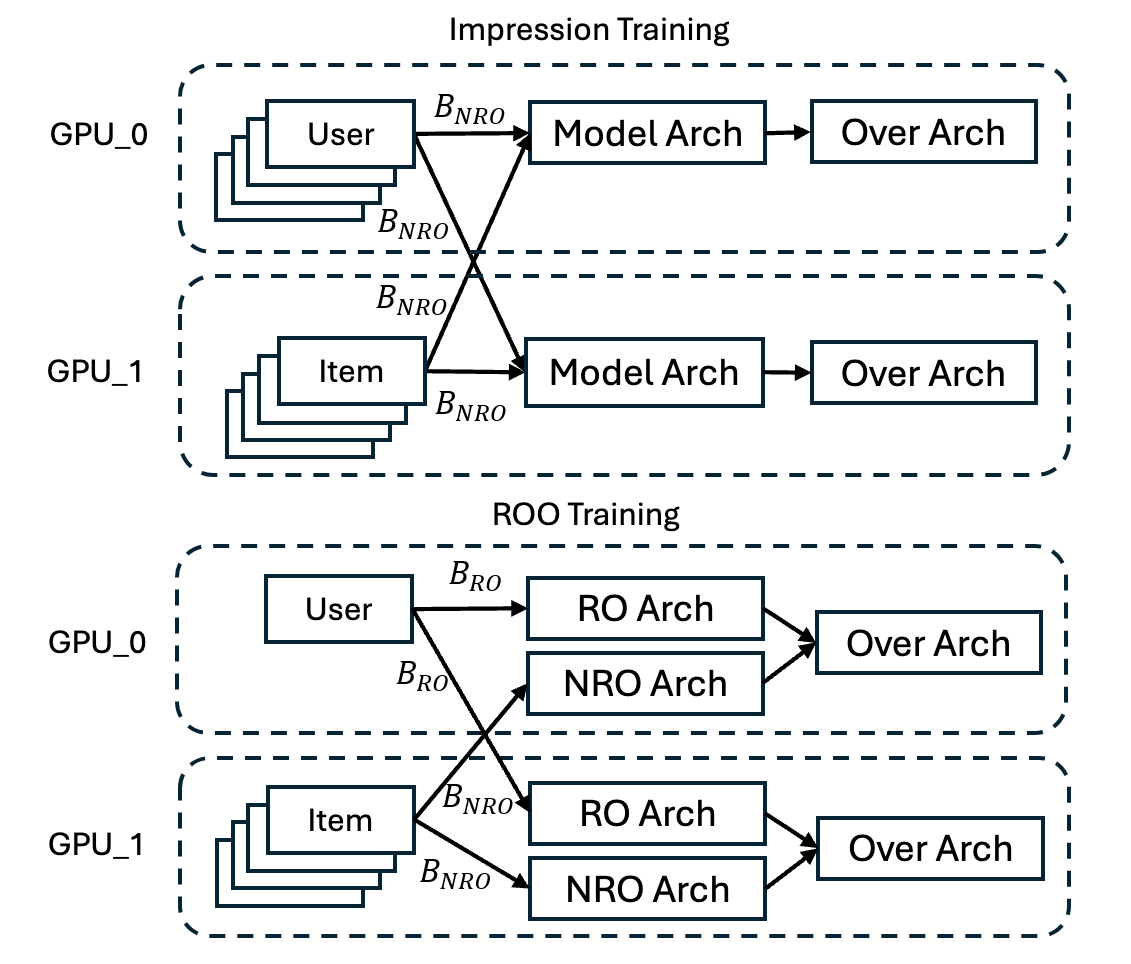}
  \caption{Compared to impression training, ROO training deduplicates RO-side embedding lookup, embedding all-to-all communication, RO-only model arch's computation FLOPs from $B_{NRO}$ batch size to $B_{RO}$ batch size. In the example, user and item embedding tables are sharded across GPU-0 and GPU-1 accordingly; $B_{NRO}$ is 4 and $B_{RO}$ is 1.}
  \Description{roo training before and after}
  \label{fig:training-paradigm}
\end{figure}

In a trainer mini-batch, the user-side RO and item-side NRO feature tensors have different batch sizes. The RO tensor batch size $B_{RO}$ is the number of ROO training samples, while the NRO tensor batch size $B_{NRO}$ is the sum of impression items across all request level samples in the trainer mini-batch. 
This difference in batch sizes can be effectively managed by leveraging TorchRec’s variable-length batch sharding, allowing the same embedding table to be shared between RO and NRO features.

The system efficiency gains primarily come from the user-side RO computation deduplication. User-side RO feature embedding lookup and associated all-to-all communication overhead are reduced to exactly once for each request-level sample in the mini-batch. Compared to impression level training, as illustrated in Figure \ref{fig:training-paradigm}, the computation and communication complexity is reduced from $O(B_{NRO})$ to $O(B_{RO})$. In our extensive experiments in DLRM systems (Section~\ref{sec:experiments}), the RO model arch reduced overall FLOPs significantly as a result of the ROO-aware training paradigm.

Traditional DLRM training and inference had different model input feature data formats that led to diverging code stacks. At the model inference side, request level features were explicitly deduplicated through coordinated inference server-client operations to save network and GPU resources. These request level optimizations were premature because they only partially mitigated the problems but added layers of complexity without providing a universal solution.

Similar to ROO training, ROO inference feature preprocessing allows DLRM inference to easily distinguish RO and NRO features within the inference request and therefore passing RO feature to recommendation models for deferred user-side feature fanout inside the model. It removes unintuitive client-side user feature broadcast and server-side deduplication. Effectively, the ROO training paradigm unifies DLRM training and inference stacks, while dramatically simplifying the DLRM system with scalability improvements and engineering cost reduction.

\section{ROO Model Architectures}
\label{sec:roo-model-archs}

Co-designing with the ROO data format and training paradigm, we investigate a novel family of DLRMs that have highly scaled-up complexity on the RO part of the model, where the cost of scaling up is largely amortized by the computation deduplication in ROO. 

\subsection{Simple RO Scale-Up: User Tower in Two-Tower Like Models}
\begin{figure*}[]
    \centering
    \begin{subfigure}[b]{0.35\linewidth}
      \centering
      \includegraphics[width=\linewidth]{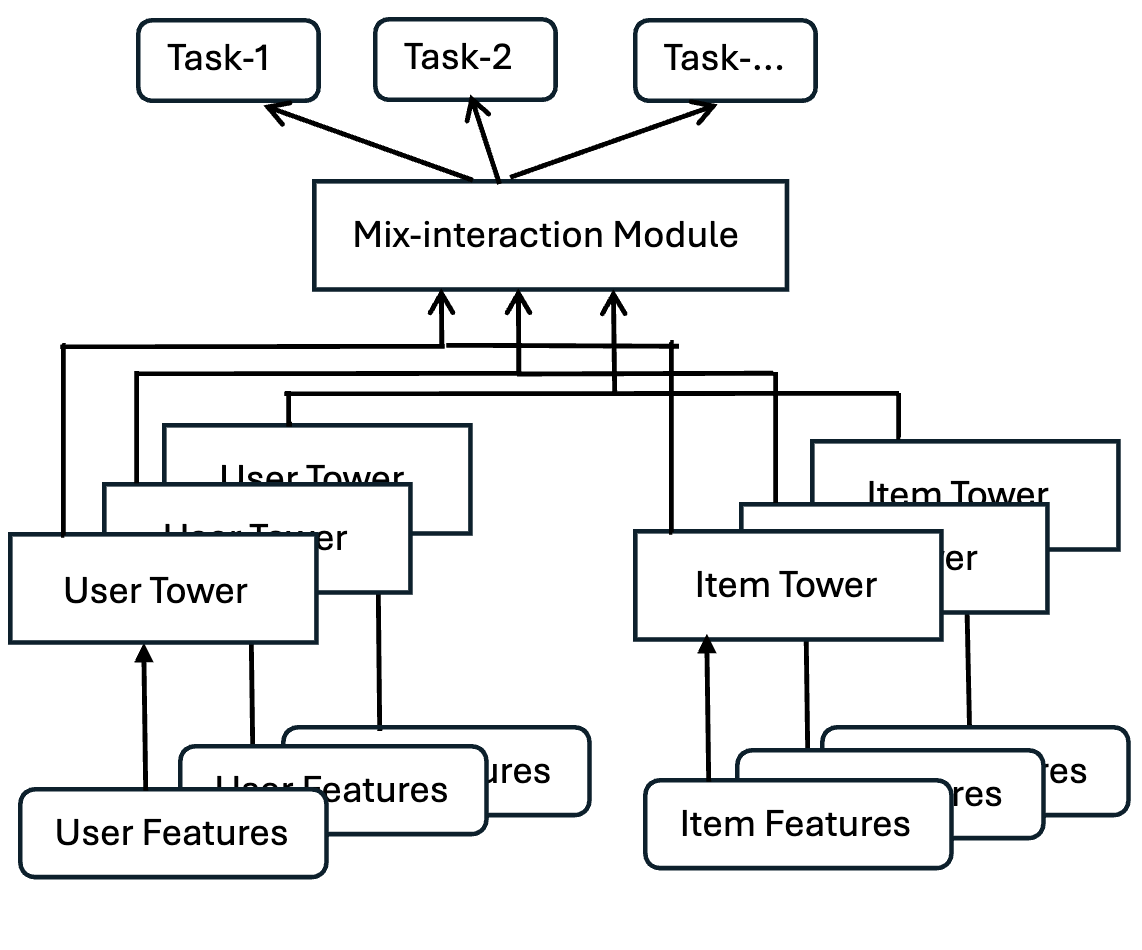}
      \caption{Impression-level training.}
      \label{fig:esr_model_arch_impression}
    \end{subfigure}%
    \hspace{0.15\linewidth}
    \begin{subfigure}[b]{.35\linewidth}
      \centering
      \includegraphics[width=\linewidth]{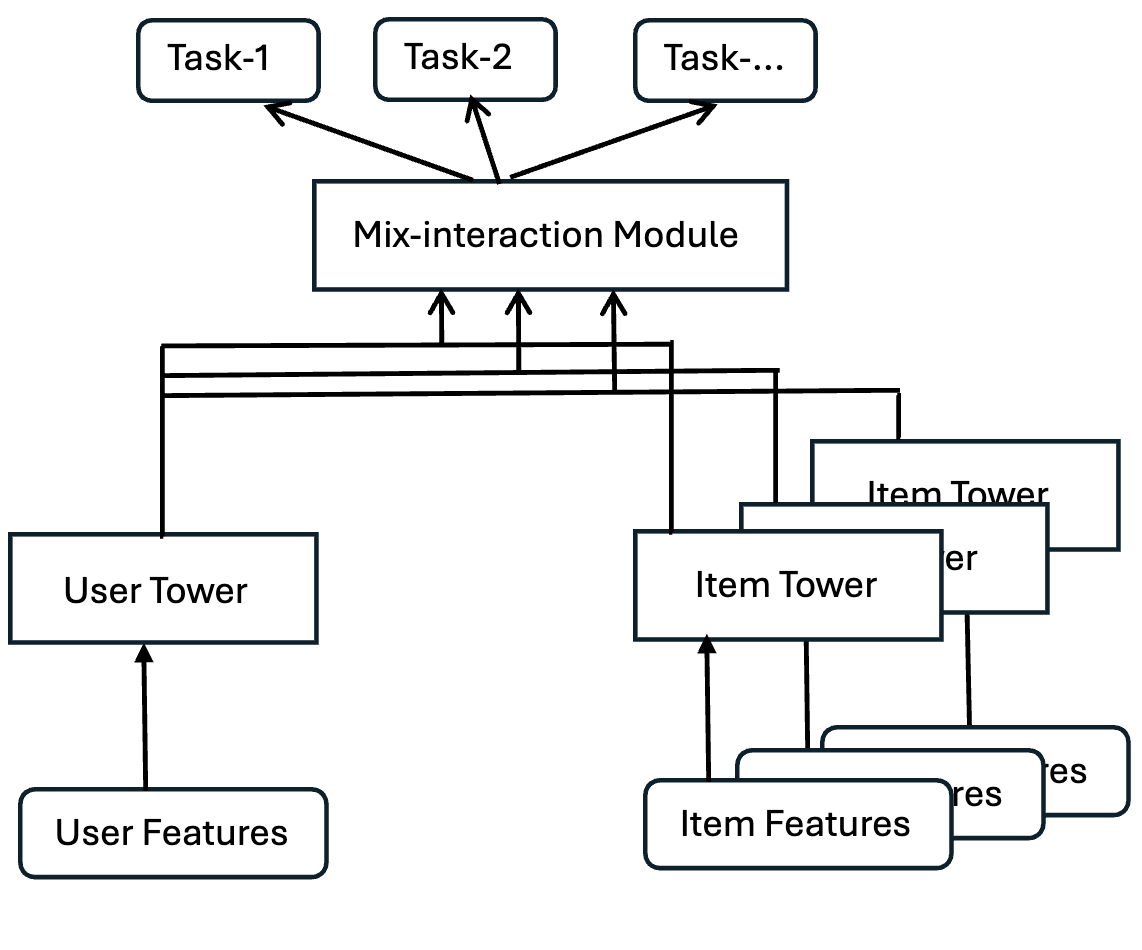}
      \caption{ROO training.}
      \label{fig:esr_model_arch_ro}
    \end{subfigure}
    \caption{The two-tower like models in the recommendation system with lightweight interaction between the two towers, e.g. retrieval models or early-stage ranking model. Figure (a) shows the models in the traditional impression-level training mode, figure (b) shows the models in the ROO training mode.}
    \label{fig:esr_model_arch}
\end{figure*}

In the funnel of typical industrial recommendation systems, the retrieval model and the early-stage ranking (ESR) model~\cite{cascade_ranking_wsdm19}, despite having different optimization goals, usually consist of a user tower that learns user representations, an item tower that learns item representations, and an optional user-item mix-interaction module fusing the user representations with item representations to capture the user-item interactions. We illustrate one such example in Figure~\ref{fig:esr_model_arch}. 

As the user tower only processes the user features and the impressions in a request have the same user features (RO features from Section~\ref{sec:roo:request-level-training-data}), the user tower computation for impressions in a request can be deduplicated, and the user representation is learned for a request and shared by these impressions. 

\subsection{RO-side Scale Up with UserArch}
Traditional deep learning recommendation models (DLRMs) have been shown to scale effectively with an increasing number of features, leading to improved performance. However, this comes at the cost of increased complexity in the interaction arch, e.g. DCNv2~\cite{wang2021dcn} and DHEN~\cite{zhang2022dhen}. In the context of recommendation systems, features can be broadly categorized into user features, candidate features, and cross features that capture the relationships between the user and candidate. Notably in recommendation systems, user features dominate the feature landscape, accounting for the majority of the features in typical ranking models for recommendation services. In late-stage ranking (LSR) models, the interaction between user and item features occurs early in the neural network, limiting the benefits of ROO training primarily to embedding table lookups or inter-GPU communications for common DLRM models. 

To fully exploit the advantages of ROO training, we designed UserArch to efficiently process and integrate user-side features. Specifically, we employ a simple cross-feature interaction architecture, Linear Compress Embedding (LCE)~\cite{zhang2022dhen,mlpmixer_neurips21,mlp4rec_ijcai22}, to compress the dimensionality of user features before they enter the post-ROO architecture. Let $X \in R^{B, d_{in}, n_{in}}$ be the tensor of user features, where $B$ is the batch size, $d_{in}$ is the embedding dimensions of each input features, and $n_{in}$ is the number of input user features. The LCE module would first compress the number of embeddings to the output number of embeddings $n_{out}$:
\begin{equation}
    f(X) = b + g(X) \odot W
\label{eq:lce_1}
\end{equation}
where $b \in R^{1, n_{out}}$, $g(X)$ will then reshape and permute $X$ to $R^{B \times d, n_{in}}$, and $W \in R^{n_{in},n_{out}}$. Then another linear layer will project all the compressed embeddings to an output dimension $d_{out}$:
\begin{equation}
    f'(f(X)) = b' + g'(f(X)) \odot W'
\end{equation}
where $b' \in R^{1, d_{out}}$, $g'(f(X))$ will the reshape and permute $f(X)$ to $R^{B \times n_{out}, d_{in}}$, and $W \in R^{d_{in},d_{out}}$. Note that similar to the sequential modeling case, the computational cost of UserArch itself can also be amortized by ROO.

This compression enables the inclusion of more user-side features without significantly increasing the size of the post-ROO architecture, thereby enhancing user-centric modeling while maintaining model efficiency. By amortizing the cost of RO-side feature generation through ROO, our UserArch design allows for the efficient incorporation of additional user-side features, ultimately improving model performance and scalability. As illustrated in Figure \ref{fig:lsr_model_arch}, in LSR models, the UserArch outputs are fed into the post-ROO architecture, adding only ROO amortized extra computation to the model before the interaction architecture of traditional DLRM for ranking.

\subsection{ROO-based Autoregressive Modeling} 
Most importantly, ROO addresses challenges associated with recent trends to move to autoregressive model training in RecSys.


ROO enables scalable autoregressive modeling, with a commonly used variant, generative recommenders (GRs)~\cite{zhai2024actions,killingbirdsstoneunifying_sigir25,huangxhs25,borisyuk2025ligr,mtgr2025,khrylchenko2025scalingrecommendertransformersbillion} as follows. Denote a sequence of $n$ past impressions in user history $RO_0, RO_1,$ $ \ldots, RO_{n-1} (RO_i \in \mathbb{X})$ ordered chronologically, their associated user actions (e.g., like or skip), and a sequence of $m$ target items in the request $NRO_0, NRO_1, \ldots, NRO_{m-1}$ ($NRO_i \in \mathbb{X}$), where $\mathbb{X}$ denotes the set of all items served in the product. With each item associated with an action feature $a$ 
and various contextual features including but not limited to surface type, timestamp, denoted as $c$, sequential modeling in standard retrieval and ranking models can be formulated as shown in Figure~\ref{fig:sequetial-modeling}.

\textbf{Amortized training cost with ROO.}
\label{sec:dlrms-to-grs-generative-training}
In a traditional impression-based setup, the total computational requirement for self-attention based sequential transduction architectures, such as Transformers~\cite{vaswani2017attention}, scales as $m*(n^2d+nd^2)$ for all impressions in the request. In our proposed ROO setup, the cost goes down to $(n+m)^2d+(n+m)d^2$, where $n$ represents the length of the user history, which is typically in the order of thousands, and m represents the request size, which is typically in the order of tens. In a moderate scenario where $n=1000, m=10, d=256$, the saving for computational cost would be $9.82x$ for the sequential encoder module.

\ysrev{
\begin{figure*}
  \centering
  \includegraphics[width=\textwidth]{
  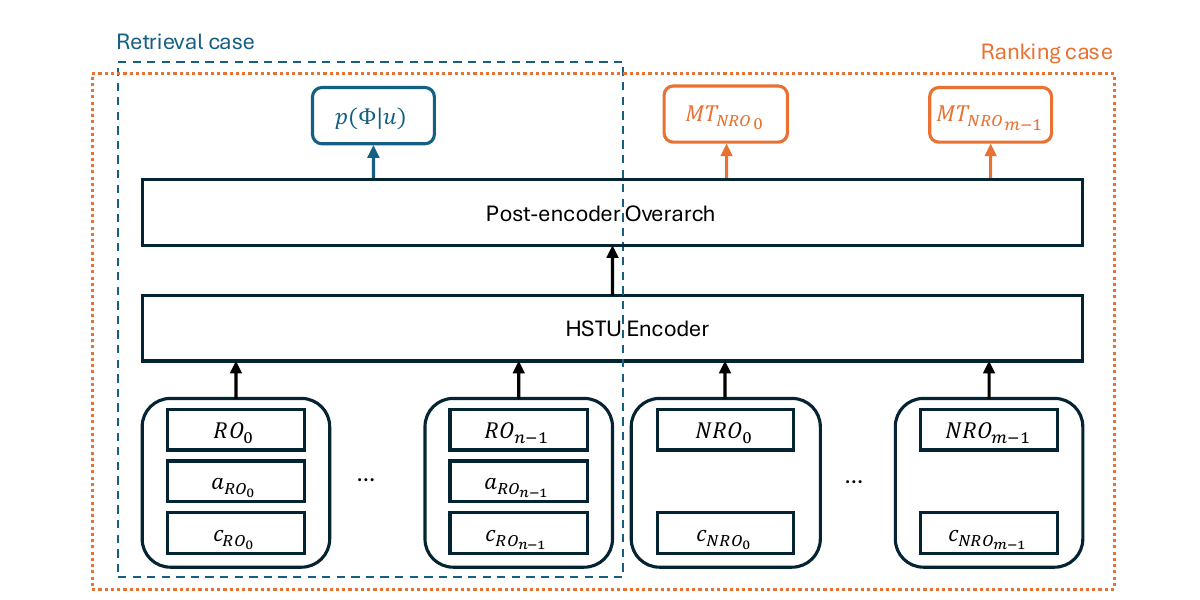}
  \caption{Sequential modeling for ROO retrieval and ranking. Note that in the retrieval case, the sequence encoder only takes user history as input. But in the ranking case, the target information is also input to the sequence encoder.}
  \Description{}
  \label{fig:sequetial-modeling}
\end{figure*}
\begin{figure} 
  \centering
  \includegraphics[width=0.7\linewidth]{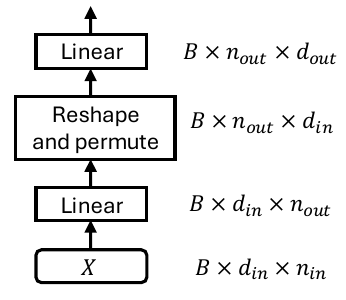}
  \caption{LCE in UserArch.}
  \Description{The UserArch illustration}
  \label{fig:user-arch}
\end{figure}
}{}
\begin{figure*}[t]
    \centering
    \begin{subfigure}[m]{.25\linewidth}
      \centering
      \vspace{5mm}
      \includegraphics[width=\linewidth]{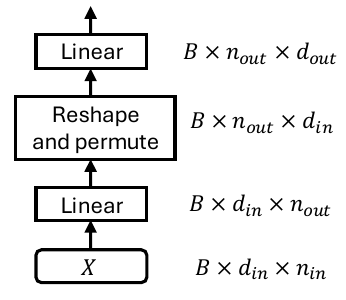}
      \vspace{2mm}
      \caption{LCE in UserArch.}
      \label{fig:user-arch}
    \end{subfigure}
    \hspace{0.05\textwidth}
    \begin{subfigure}[m]{0.65\linewidth}
      \centering
      \includegraphics[width=\textwidth]{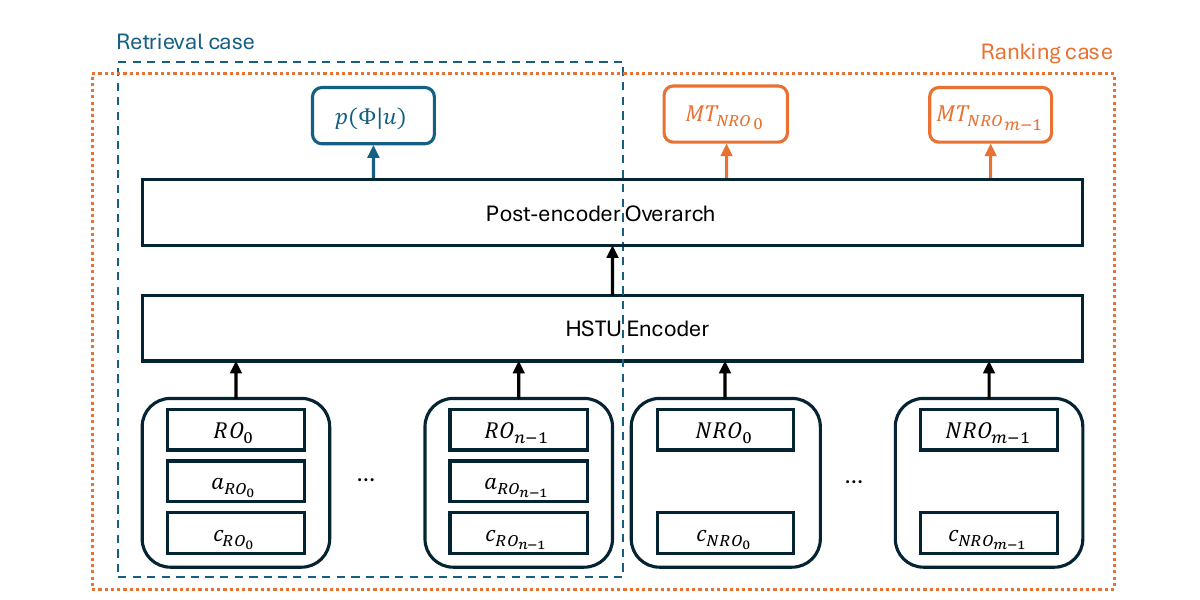}
      \caption{Sequential modeling for ROO retrieval and ranking. Note that in the retrieval case, the sequence encoder only takes user history as input, but in the ranking case, the target information is also input to the sequence encoder.}
      \label{fig:sequetial-modeling}
    \end{subfigure}%
    \vspace{-.5em}
    \caption{Sequential modeling and UserArch.} 
    \vspace{-2mm}
\end{figure*}
\textbf{HSTU.} For autoregressive modeling-related experiments in this paper, we apply HSTU~\cite{zhai2024actions} -- a variant of transformer designed for industrial-scale recommendation systems with large, non-stationary vocabularies -- over the user sequence for aggregating and capturing nuanced interests and interactions. We scale up HSTU leveraging the opportunity in deduplicating computation that depends only on the RO part of the ROO data.

\section{Experiments}
\label{sec:experiments}

\subsection{End-to-End System Efficiency Savings}
ROO training data captures all impressions of the same request in one single request-level training sample and hereby eliminates duplications at the fundamental data level. Through data and system co-design, we eliminated computational redundancy throughout training data pipeline, intra-datacenter network transfer, feature preprocessing, GPU training, and model inference. As a result, ROO training achieved multi-fold efficiency benefits. First we benchmarked the efficiency savings from ROO through a range of experiments in the retrieval, early-stage ranking (ESR), and late-stage ranking (LSR) models from three billion-user scale recommendation products, hereinafter referred to as product-A, -B, -C, and -D.


ROO data allows logging more training samples and avoids downsampling with the same storage space. As shown in Table \ref{tab:traing-sample-volume-increases},  we observed 43\% to 150\% increases in training sample volumes across a few recommendation products. The storage space saving depends heavily on the proportion of user (request) side features and the number of training examples within a request. Heterogeneous user behaviors across different recommendation products leads to different training example distributions. In the extreme case, if a request contains only one training example, ROO data effectively reverts to impression level data, and no de-duplication is possible.


\begin{table}[]
    \centering
    \begin{tabular}{cc}
        \toprule
         & Training samples increase \\
         \midrule
        Product-A & 43\% \\
        Product-B & 150\% \\
        Product-C & 100\% \\
        \bottomrule
    \end{tabular}
    \vspace{.3em}
    \caption{Training sample volume increase across different recommendation products.}
    \label{tab:traing-sample-volume-increases}
\end{table}


ROO training significantly boosts training throughput, enabling the exploration of previously compute-prohibitive modeling technologies like UserArch and GRs/HSTU. We benchmarked ROO and impression-level training throughput across multiple models and recommendation stages. In all experiments, the training data originated from the same industrial recommendation source, and the neural network architectures remained consistent. The only controlled variations were the ROO data, preprocessing, and training pipeline.

As shown in Table~\ref{tab:qps}, our proposed ROO techniques achieved substantial increases in training throughput: 220\% to 570\% in retrieval models, 125\% to 266\% in early-stage ranking models, and 32\% to 100\% in late-stage ranking models. In practice, these throughput improvements are also influenced by the percentage of compute and memory that can be de-duplicated within individual models. Two-tower models with a heavy user tower, such as retrieval and ESR models., benefit more from ROO than the LSR models.

In recommendation product-D surface, we compared a RecD-trained ESR model with ROO-trained model where ROO training showed 50\% more training throughput.



\begin{table}[]
    \centering
    \begin{tabular}{rr}
        \toprule
        Model & Training throughput increase \\
        \midrule
        LSR Product-A & 50\% \\
        ESR Product-A & 125\% \\
        \midrule
        LSR Product-B & 33\% \\
        ESR Product-B & 266\% \\
        Retrieval Product-B & 570\% \\
        \midrule
        LSR Product-C & 32\% \\
        ESR Product-C & 150\% \\
        Retrieval Product-C & 220\% \\
        \bottomrule
    \end{tabular}
    \vspace{.3em}
    \caption{Training throughput increase (in percentage) on multiple models across the recommendation stages.}
    \label{tab:qps}
\end{table}

\subsection{Model Quality Improvements}

We next conduct experiments to validate the practicality of the ROO paradigm and the related model architecture innovations in production environments. We also provide detailed ablation studies and online A/B test results to showcase the impact from ROO in production.

In the following subsections that discuss offline studies of early-stage and late-stage ranking models (ESR, LSR), we use normalized entropy (NE)~\cite{he2014practical} as evaluation metrics, which indicates better accuracy if the value is reduced. We use Recall@K for retrieval models, where better models have better recall metrics.


\subsection{Retrieval and Early-Stage Ranking Models}

\begin{table*}
  \caption{Relative FLOPs and offline/online metric comparisons (Recall@100 for the main C-/E-Task)
  for retrieval models.}\label{tab:retrieval-recall-offline}
  \vspace{-.5em}
  \begin{tabular}{cccccc}
    \toprule
    {Model} & & \thead{\normalsize FLOPs per \\ \normalsize example} & \thead{\normalsize C-Task \\ 
 \normalsize Recall@100} & \thead{\normalsize E-Task \\ \normalsize Recall@100} & \thead{\normalsize Topline Consumption} \\
    \midrule
   \multirow{3}{*}{Retrieval} & HSTU (Impression) & 6.8x & +3.53\% & +2.35\% & --\\
    & HSTU (ROO)  & 0.99x & +3.87\% & +2.41\% & +4.77\%\\
    & Baseline  & 1.0x & +0.0\% & +0.0\% & +0.0\% \\
  \bottomrule
\end{tabular}
\end{table*}

Through reducing training costs per item example by deduplicating user tower computation costs across items in a logged request, ROO has enabled the development of more complex user architectures for retrieval and early-stage ranking models.

\subsubsection{Retrieval Models}

Table~\ref{tab:retrieval-recall-offline} illustrates the value of ROO optimization for model scale-up. Here we apply a dual encoder-based retrieval model~\cite{ytdnn_goog_recsys16,fbsearch_arxiv2020,ccs_kdd24} 
to illustrate relative training cost and model quality comparisons when the same model is migrated to ROO, although our approach can be easily extended to recent work including multi-embeddings~\citep{mind_2019}, generative retrieval~\citep{dr_cikm21,tiger_rajput2023recommender}, and learned similarities~\cite{zhai2023revisiting,rails_www25}. 

Without the ROO data format, model scaling can contribute to improved offline results at the expense of 6.8x more computation costs per training example. With ROO, the same improvements in offline recall can be attained with no increase in training cost per item example as we now effectively amortize computational cost across multiple impressions. We collected online A/B test metrics from a large-scale experiment on product-B. 
We observed a 4.77\% increase in the main online consumption metric after launching the HSTU architecture on ROO.

\begin{table*}[]
  \caption{Normalized entropy (NE) improvement in offline experiments of ranking models from multiple product surfaces and online A/B test metrics improvements after the models are deployed to production. The significant level of each metric across different models are marked in separate rows. 
  In the LSR model on product-A, UserArch and HSTU are tested and deployed sequentially so that their metrics are listed incrementally. E.g. the two modules brings 0.41\% and 0.53\% NE improvement (reduction) on E-task respectively.}
  \label{tab:ranking-metrics}
  \begin{tabular}{rlccccc}
    \toprule
    Model & & E-Task NE&C-Task NE&Topline&Consumption& Engagement\\
    \midrule
    LSR on product-A & UserArch & -0.41\%   & -0.42\%   & +0.04\%  & +0.08\%  & -\\
    & \quad  +HSTU & -0.53\%   & -0.57\%   & +0.12\%  & +0.18\%  & +4.8\% \\
    \it Significant delta on product-A &  & 0.1\% & 0.1\%   & 0.03\%  & 0.06\% & 0.63\% \\
    \midrule
    ESR on product-B &  HSTU & -0.31\%  & -0.32\%  & +0.04\%  & +2.43\%  & +4.9\%  \\
    LSR on product-B &   UserArch+HSTU & -0.86\%  & -0.78\%  & +0.07\%  & +5.56\%  & +7.0\%  \\
    \it Significant delta on product-B &  & 0.1\% & 0.1\%   & 0.03\%  & 0.24\% & 0.63\% \\
    \midrule
    LSR on product-C &  HSTU & -0.81\%  & -0.79\%  & +0.25\%  & +1.1\% & +2.9\% \\
    \it Significant delta on product-C &  & 0.1\% & 0.1\%   & 0.03\%  & 0.04\% & 0.19\% \\
    \midrule
    LSR on product-D &   UserArch+HSTU & -0.60\%  & -0.64\%  & +0.04\%  & +2.51\%  & +6.9\%  \\
    \it Significant delta on product-D &  & 0.1\% & 0.1\%   & 0.03\%  & 0.24\% & 0.63\% \\
  \bottomrule
\end{tabular}
\end{table*}


\subsubsection{Early-stage Ranking (ESR) Models}
With the ROO paradigm, we further integrated HSTU into the user tower of the ESR models. As shown in Table~\ref{tab:ranking-metrics}, HSTU improved the recommendation quality as measured by both offline NE and post launch topline impact.  Specifically, this design leads to significant boost of the main consumption metric by 2.43\% and the main topline metric by 0.04\% to product-B, where 0.24\% and 0.03\% are significant, respectively.


\subsection{Late-Stage Ranking (LSR) Models}

We finally conduct extensive experiments on four LSR models across three recommendation products.

\textbf{LSR on product-A.}
In this experiment, we conducted an offline ablation analysis on a late-stage ranking model for product-A. We incorporated two ROO architectures into the model: UserArch and HSTU. 
On this model, we deployed UserArch to production first, followed by HSTU, which was tested and rolled out subsequently. From both offline and online experiments, HSTU added notable incremental improvement to the model on top of UserArch. Table \ref{tab:ranking-metrics} summarizes the offline evaluation NE improvement and post-launch online metrics lift. Note that a 0.1\% NE improvement on both engagement and consumption tasks is considered substantial for this particular model, and 0.03\% lift in the topline metric, 0.63\% lift in engagement metric, and 0.06\% lift in consumption metric are deemed significant for this product.



\textbf{LSR on product-B.}
We conducted a evaluation of UserArch and HSTU on an LSR model within the recommendation service for product-B. In this product, 0.03\% lift in the topline metric, 0.63\% lift in engagement metric, and 0.24\% lift in consumption metric is considered as significant. Our offline and online experiments revealed that these ROO-related technologies yielded notable improvements in NE and post-launch online metrics in both models as shown in Table \ref{tab:ranking-metrics}.

\textbf{LSR on product-C.}
We tested HSTU on a LSR model for product-C, yielding significant improvements in NE. Specifically, we observed 0.81\% and 0.79\% NE reductions on engagement and consumption tasks, respectively, which are larger than the 0.1\% significant level. In online A/B test, we observed substantial gains in online metrics, including a 0.25\% lift in the primary topline metric and a 2.85\% increase in engagement metric as shown in Table \ref{tab:ranking-metrics}. Normally, 0.03\% lift in topline metric, 0.19\% in engagement metric and 0.04\% in consumption metric are considered significant for this product.  \ysrev{Note: can put numbers here in a table too. It's hard to read in text.}{} 

\textbf{LSR on product-D.}
On product-D, we tested UserArch and HSTU on an LSR model. We have observed distinct improvements in both offline and online experiments exceeding the significant level as shown in Table \ref{tab:ranking-metrics}. Particularly, in offline experiment, 0.60\% and 0.64\% NE reductions on engagement and consumption tasks were observed. In online A/B test, the topline, consumption, and engagement metrics were lifted by 0.04\%, 2.51\%, and 6.9\% respectively.
\section{Related Work}
\label{sec:related-work}
\textbf{Data de-duplication.} 
Normalization~\cite{codd1970relational} is a process to facilitate redundancy removal and data integrity checking in relational databases. The impression-level training sample schema (Table~\ref{tab:impression_sample}) violates the second normal form (i.e. 2NF) \cite{codd1972further}, which prohibits a non-key column being a fact about only a subset of the key columns~\cite{kent1983simple}. To achieve the 2NF and data de-duplication on the training sample schema, the RO side features should be extracted out to another table with request id being the key. Request-level schema (Table~\ref{tab:request_level_sample}), in contrast, exploited the RO feature equivalence within a request by grouping training samples from the same request into one sample. Despite the aforementioned differences, the timelessness idea of normalization inspired the need to remove data redundancy and the design of ROO schema.

In the field of DLRM, data de-duplication has also been investigated~\cite{zhou2022serving, gai2017learning, you2024exploiting}. In particular, part of our work in the inference system is reminiscent of the work done by You et al.~\cite{you2024exploiting}, which described the duplication as “single-user-multiple-candidates” and introduced a “structured feature” to exploit the de-duplication to reduce cost during inference time. However, these studies generally focused on one part of the system (e.g. inference time). Our work provides a comprehensive analysis of feature value de-duplication throughout the end-to-end data, training and inference pipeline. 

RecD \cite{zhao2023recd} addressed feature duplication through data pipeline optimizations, grouping impression samples to maximize columnar storage compression and using an inverse index for reconstruction during training. Our approach differs by offering a holistic system-wide efficiency across storage, training, and inference. The ROO data schema deduplicates features at the source, avoiding data pipeline infrastructure changes or increased landing latency. Its preprocessing integrates seamlessly with common tensor structures (e.g., KeyedJaggedTensors in TorchRec \cite{ivchenko2022torchrec}), eliminating the need for custom structures like InverseKeyedJaggedTensors and minimizing model rewrite efforts. More implementation details on ROO data are provided in the appendix \ref{app:request-level-join}.

\textbf{Efficient scaling of DLRM}.
To scale DLRMs, an effective line of thinking is to examine the recommendation life-cycle from an end-to-end point of view, seeking model-system co-design and global optimization opportunities. Efficient application of various hardware (e.g. accelerators, scheduler, SSDs) in different components of the end-to-end recommendation system has been studied in \cite{mudigere2022software,gupta2020deeprecsys,wilkening2021recssd}. In particular, some work adapted the ML architectures deliberately to enable more efficient usages of the accelerator resources \cite{xu2022rethinking,zhai2023revisiting}. In addition, leveraging offline/batch compute that has access to more compute and memory resources is also observed in many systems \cite{pancha2022pinnerformer,pi2019practice}.

Optimizing the high computation cost in attention-based methods for long user behavior sequence modeling has attracted a lot of interests. Many work studied cost effective techniques to select sub-user behavior sequences that are relevant to candidates \cite{zhou2019deep,cao2022sampling,feng2024context}. Following up on DIN's \cite{zhou2019deep} target attention work, Pi et al.~\cite{pi2020search} employed a two-stage approach to select target aware sub-sequences in the general search stage and perform finer grained modeling in the exact search stage. On top of the new two-stage paradigm, Chang, et al.~\cite{chang2023twin} and the more recent work by Lv, et al.~\cite{lv2025marm} adopted a similar two-stage architecture but supported by pre-computing and caching strategies for online inference, which enhanced system performance. Si et al.~\cite{si2024twin}'s work pre-computed a hierarchical clustering of the long user sequence offline, and allowed subsequence selection based on clusters during inference. Unlike pairwise attention approaches, HSTU \cite{zhai2024actions} leveraged efficiently-designed self-attention sequential transduction units to capture the complex interactions between target and history sequences.

Orthogonal to the methods above, the ROO paradigm provides a generic solution focused on systematic de-duplication and amortization, highlighting the importance of model-system co-design to optimize data, training, and inference in the DLRM context.

\section{Conclusion}
\label{sec:conclusions}
In this paper, we introduced the Request-Only Optimization (ROO) training and modeling paradigm, a holistic approach that co-designs data, system, and model architectures to unlock unprecedented efficiency and scalability in deep learning recommendation models. By addressing the feature duplication problem at its source through our innovative ROO training data schema design, we achieved significant gains in system efficiency and unified request-only optimization across training and inference.
Our results demonstrate substantial improvements in training throughput, with 2x to 6x efficiency gains across retrieval and ranking models. Moreover, our ROO model architectures have consistently delivered enhanced offline and online A/B test metrics, driving impact across multiple applications serving billions of users worldwide.
The widespread adoption of the ROO training and modeling paradigm in major DLRM products processing billion-scale user traffic is a testament to its effectiveness and versatility. We hope that the lessons we share in this paper are useful to industry practitioners seeking to optimize their own deep learning recommendation systems.

\section{Acknowledgment}
This work would not be possible without work from the following contributors (alphabetical order): Bugra Akyildiz, Yuanfei Bi, Duo Chen, Li Chen, Xi Chen, Xianjie Chen, Rex Cheung, Ted Cui, Patrick Cullen, Delia David, Fei Ding, Rajeev Dixit, Md Mustafijur Rahman Faysal, Bo Feng, Liz Guo, Luyi Guo, Daisy Shi He, Michael He, Austin Jenkins, Jie Jiao, Basri Kahveci, Manos Karpathiotakis, Keyu Lai, Dai Li, Shen Li, Hao Lin,  Weiran Liu, Yinghai Lu, Michael Ly, Matt Ma, Yun Mao, Franco Mo, Min Ni, Jongsoo Park, Jing Qian, Alex Singh, Jason Song, Bingjun Sun, Ruohan Sun, Yimin Tan, Alex Vu, Shanyue Wang, Wenyu Wang, Zellux Wang, Ziqi Wang, Evan Welch, Yue Weng, Yanhong Wu, Ke Xu, Zheng Yan, Zimeng Yang, James Zhang, Jingjing Zhang, Lu Zhang, Qunshu Zhang, Vivian Zhang, Wei Zhang, Yue Zhang, Zhiyun Zhang, Xin Zhuang. 

\newpage

\bibliographystyle{ACM-Reference-Format}
\balance
\bibliography{sample-base}


\begin{thebibliography}{49}


\ifx \showCODEN    \undefined \def \showCODEN     #1{\unskip}     \fi
\ifx \showISBNx    \undefined \def \showISBNx     #1{\unskip}     \fi
\ifx \showISBNxiii \undefined \def \showISBNxiii  #1{\unskip}     \fi
\ifx \showISSN     \undefined \def \showISSN      #1{\unskip}     \fi
\ifx \showLCCN     \undefined \def \showLCCN      #1{\unskip}     \fi
\ifx \shownote     \undefined \def \shownote      #1{#1}          \fi
\ifx \showarticletitle \undefined \def \showarticletitle #1{#1}   \fi
\ifx \showURL      \undefined \def \showURL       {\relax}        \fi
\providecommand\bibfield[2]{#2}
\providecommand\bibinfo[2]{#2}
\providecommand\natexlab[1]{#1}
\providecommand\showeprint[2][]{arXiv:#2}

\bibitem[Borisyuk et~al\mbox{.}(2025)]%
        {borisyuk2025ligr}
\bibfield{author}{\bibinfo{person}{Fedor Borisyuk}, \bibinfo{person}{Lars Hertel}, \bibinfo{person}{Ganesh Parameswaran}, \bibinfo{person}{Gaurav Srivastava}, \bibinfo{person}{Sudarshan~Srinivasa Ramanujam}, \bibinfo{person}{Borja Ocejo}, \bibinfo{person}{Peng Du}, \bibinfo{person}{Andrei Akterskii}, \bibinfo{person}{Neil Daftary}, \bibinfo{person}{Shao Tang}, \bibinfo{person}{Daqi Sun}, \bibinfo{person}{Qiang~Charles Xiao}, \bibinfo{person}{Deepesh Nathani}, \bibinfo{person}{Mohit Kothari}, \bibinfo{person}{Yun Dai}, {and} \bibinfo{person}{Aman Gupta}.} \bibinfo{year}{2025}\natexlab{}.
\newblock \bibinfo{title}{From Features to Transformers: Redefining Ranking for Scalable Impact}.
\newblock
\showeprint[arxiv]{2502.03417}~[cs.LG]
\urldef\tempurl%
\url{https://arxiv.org/abs/2502.03417}
\showURL{%
\tempurl}


\bibitem[Cao et~al\mbox{.}(2022)]%
        {cao2022sampling}
\bibfield{author}{\bibinfo{person}{Yue Cao}, \bibinfo{person}{Xiaojiang Zhou}, \bibinfo{person}{Jiaqi Feng}, \bibinfo{person}{Peihao Huang}, \bibinfo{person}{Yao Xiao}, \bibinfo{person}{Dayao Chen}, {and} \bibinfo{person}{Sheng Chen}.} \bibinfo{year}{2022}\natexlab{}.
\newblock \showarticletitle{Sampling is all you need on modeling long-term user behaviors for CTR prediction}. In \bibinfo{booktitle}{\emph{Proceedings of the 31st ACM International Conference on Information \& Knowledge Management}}. \bibinfo{pages}{2974--2983}.
\newblock


\bibitem[Chang et~al\mbox{.}(2023)]%
        {chang2023twin}
\bibfield{author}{\bibinfo{person}{Jianxin Chang}, \bibinfo{person}{Chenbin Zhang}, \bibinfo{person}{Zhiyi Fu}, \bibinfo{person}{Xiaoxue Zang}, \bibinfo{person}{Lin Guan}, \bibinfo{person}{Jing Lu}, \bibinfo{person}{Yiqun Hui}, \bibinfo{person}{Dewei Leng}, \bibinfo{person}{Yanan Niu}, \bibinfo{person}{Yang Song}, {et~al\mbox{.}}} \bibinfo{year}{2023}\natexlab{}.
\newblock \showarticletitle{TWIN: TWo-stage interest network for lifelong user behavior modeling in CTR prediction at kuaishou}. In \bibinfo{booktitle}{\emph{Proceedings of the 29th ACM SIGKDD Conference on Knowledge Discovery and Data Mining}}. \bibinfo{pages}{3785--3794}.
\newblock


\bibitem[Codd(1970)]%
        {codd1970relational}
\bibfield{author}{\bibinfo{person}{Edgar~F Codd}.} \bibinfo{year}{1970}\natexlab{}.
\newblock \showarticletitle{A relational model of data for large shared data banks}.
\newblock \bibinfo{journal}{\emph{Commun. ACM}} \bibinfo{volume}{13}, \bibinfo{number}{6} (\bibinfo{year}{1970}), \bibinfo{pages}{377--387}.
\newblock


\bibitem[Codd(1972)]%
        {codd1972further}
\bibfield{author}{\bibinfo{person}{Edgar~F Codd}.} \bibinfo{year}{1972}\natexlab{}.
\newblock \showarticletitle{Further normalization of the data base relational model}.
\newblock \bibinfo{journal}{\emph{Data base systems}}  \bibinfo{volume}{6} (\bibinfo{year}{1972}), \bibinfo{pages}{33--64}.
\newblock


\bibitem[Covington et~al\mbox{.}(2016)]%
        {ytdnn_goog_recsys16}
\bibfield{author}{\bibinfo{person}{Paul Covington}, \bibinfo{person}{Jay Adams}, {and} \bibinfo{person}{Emre Sargin}.} \bibinfo{year}{2016}\natexlab{}.
\newblock \showarticletitle{Deep Neural Networks for YouTube Recommendations}. In \bibinfo{booktitle}{\emph{Proceedings of the 10th ACM Conference on Recommender Systems}} \emph{(\bibinfo{series}{RecSys '16})}. \bibinfo{pages}{191–198}.
\newblock
\showISBNx{9781450340359}


\bibitem[Ding and Zhai(2025)]%
        {rails_www25}
\bibfield{author}{\bibinfo{person}{Bailu Ding} {and} \bibinfo{person}{Jiaqi Zhai}.} \bibinfo{year}{2025}\natexlab{}.
\newblock \showarticletitle{Retrieval with Learned Similarities}. In \bibinfo{booktitle}{\emph{Proceedings of the ACM on Web Conference 2025}} (Sydney NSW, Australia) \emph{(\bibinfo{series}{WWW '25})}. \bibinfo{publisher}{Association for Computing Machinery}, \bibinfo{address}{New York, NY, USA}, \bibinfo{pages}{1626–1637}.
\newblock
\showISBNx{9798400712746}
\href{https://doi.org/10.1145/3696410.3714822}{doi:\nolinkurl{10.1145/3696410.3714822}}


\bibitem[Feng et~al\mbox{.}(2024)]%
        {feng2024context}
\bibfield{author}{\bibinfo{person}{Zhichao Feng}, \bibinfo{person}{JunJie Xie}, \bibinfo{person}{Kaiyuan Li}, \bibinfo{person}{Yu Qin}, \bibinfo{person}{Pengfei Wang}, \bibinfo{person}{Qianzhong Li}, \bibinfo{person}{Bin Yin}, \bibinfo{person}{Xiang Li}, \bibinfo{person}{Wei Lin}, {and} \bibinfo{person}{Shangguang Wang}.} \bibinfo{year}{2024}\natexlab{}.
\newblock \showarticletitle{Context-based Fast Recommendation Strategy for Long User Behavior Sequence in Meituan Waimai}. In \bibinfo{booktitle}{\emph{Companion Proceedings of the ACM on Web Conference 2024}}. \bibinfo{pages}{355--363}.
\newblock


\bibitem[Gai et~al\mbox{.}(2017)]%
        {gai2017learning}
\bibfield{author}{\bibinfo{person}{Kun Gai}, \bibinfo{person}{Xiaoqiang Zhu}, \bibinfo{person}{Han Li}, \bibinfo{person}{Kai Liu}, {and} \bibinfo{person}{Zhe Wang}.} \bibinfo{year}{2017}\natexlab{}.
\newblock \showarticletitle{Learning piece-wise linear models from large scale data for ad click prediction}.
\newblock \bibinfo{journal}{\emph{arXiv preprint arXiv:1704.05194}} (\bibinfo{year}{2017}).
\newblock


\bibitem[Gallagher et~al\mbox{.}(2019)]%
        {cascade_ranking_wsdm19}
\bibfield{author}{\bibinfo{person}{Luke Gallagher}, \bibinfo{person}{Ruey-Cheng Chen}, \bibinfo{person}{Roi Blanco}, {and} \bibinfo{person}{J.~Shane Culpepper}.} \bibinfo{year}{2019}\natexlab{}.
\newblock \showarticletitle{Joint Optimization of Cascade Ranking Models}. In \bibinfo{booktitle}{\emph{Proceedings of the Twelfth ACM International Conference on Web Search and Data Mining}} (Melbourne VIC, Australia) \emph{(\bibinfo{series}{WSDM '19})}. \bibinfo{publisher}{Association for Computing Machinery}, \bibinfo{address}{New York, NY, USA}, \bibinfo{pages}{15–23}.
\newblock
\showISBNx{9781450359405}
\href{https://doi.org/10.1145/3289600.3290986}{doi:\nolinkurl{10.1145/3289600.3290986}}


\bibitem[Gangidi et~al\mbox{.}(2024)]%
        {sigcomm24_rdma_meta}
\bibfield{author}{\bibinfo{person}{Adithya Gangidi}, \bibinfo{person}{Rui Miao}, \bibinfo{person}{Shengbao Zheng}, \bibinfo{person}{Sai~Jayesh Bondu}, \bibinfo{person}{Guilherme Goes}, \bibinfo{person}{Hany Morsy}, \bibinfo{person}{Rohit Puri}, \bibinfo{person}{Mohammad Riftadi}, \bibinfo{person}{Ashmitha~Jeevaraj Shetty}, \bibinfo{person}{Jingyi Yang}, \bibinfo{person}{Shuqiang Zhang}, \bibinfo{person}{Mikel~Jimenez Fernandez}, \bibinfo{person}{Shashidhar Gandham}, {and} \bibinfo{person}{Hongyi Zeng}.} \bibinfo{year}{2024}\natexlab{}.
\newblock \showarticletitle{RDMA over Ethernet for Distributed Training at Meta Scale}. In \bibinfo{booktitle}{\emph{Proceedings of the ACM SIGCOMM 2024 Conference}} (Sydney, NSW, Australia) \emph{(\bibinfo{series}{ACM SIGCOMM '24})}. \bibinfo{publisher}{Association for Computing Machinery}, \bibinfo{address}{New York, NY, USA}, \bibinfo{pages}{57–70}.
\newblock
\showISBNx{9798400706141}
\href{https://doi.org/10.1145/3651890.3672233}{doi:\nolinkurl{10.1145/3651890.3672233}}


\bibitem[Gao et~al\mbox{.}(2021)]%
        {dr_cikm21}
\bibfield{author}{\bibinfo{person}{Weihao Gao}, \bibinfo{person}{Xiangjun Fan}, \bibinfo{person}{Chong Wang}, \bibinfo{person}{Jiankai Sun}, \bibinfo{person}{Kai Jia}, \bibinfo{person}{Wenzi Xiao}, \bibinfo{person}{Ruofan Ding}, \bibinfo{person}{Xingyan Bin}, \bibinfo{person}{Hui Yang}, {and} \bibinfo{person}{Xiaobing Liu}.} \bibinfo{year}{2021}\natexlab{}.
\newblock \showarticletitle{Learning An End-to-End Structure for Retrieval in Large-Scale Recommendations}. In \bibinfo{booktitle}{\emph{Proceedings of the 30th ACM International Conference on Information \& Knowledge Management}} (Virtual Event, Queensland, Australia) \emph{(\bibinfo{series}{CIKM '21})}. \bibinfo{publisher}{Association for Computing Machinery}, \bibinfo{address}{New York, NY, USA}, \bibinfo{pages}{524–533}.
\newblock
\showISBNx{9781450384469}
\href{https://doi.org/10.1145/3459637.3482362}{doi:\nolinkurl{10.1145/3459637.3482362}}


\bibitem[Gupta et~al\mbox{.}(2020a)]%
        {gupta2020deeprecsys}
\bibfield{author}{\bibinfo{person}{Udit Gupta}, \bibinfo{person}{Samuel Hsia}, \bibinfo{person}{Vikram Saraph}, \bibinfo{person}{Xiaodong Wang}, \bibinfo{person}{Brandon Reagen}, \bibinfo{person}{Gu-Yeon Wei}, \bibinfo{person}{Hsien-Hsin~S Lee}, \bibinfo{person}{David Brooks}, {and} \bibinfo{person}{Carole-Jean Wu}.} \bibinfo{year}{2020}\natexlab{a}.
\newblock \showarticletitle{Deeprecsys: A system for optimizing end-to-end at-scale neural recommendation inference}. In \bibinfo{booktitle}{\emph{2020 ACM/IEEE 47th Annual International Symposium on Computer Architecture (ISCA)}}. IEEE, \bibinfo{pages}{982--995}.
\newblock


\bibitem[Gupta et~al\mbox{.}(2020b)]%
        {gupta2020architectural}
\bibfield{author}{\bibinfo{person}{Udit Gupta}, \bibinfo{person}{Carole-Jean Wu}, \bibinfo{person}{Xiaodong Wang}, \bibinfo{person}{Maxim Naumov}, \bibinfo{person}{Brandon Reagen}, \bibinfo{person}{David Brooks}, \bibinfo{person}{Bradford Cottel}, \bibinfo{person}{Kim Hazelwood}, \bibinfo{person}{Mark Hempstead}, \bibinfo{person}{Bill Jia}, {et~al\mbox{.}}} \bibinfo{year}{2020}\natexlab{b}.
\newblock \showarticletitle{The architectural implications of facebook's dnn-based personalized recommendation}. In \bibinfo{booktitle}{\emph{2020 IEEE International Symposium on High Performance Computer Architecture (HPCA)}}. IEEE, \bibinfo{pages}{488--501}.
\newblock


\bibitem[Han et~al\mbox{.}(2025)]%
        {mtgr2025}
\bibfield{author}{\bibinfo{person}{Ruidong Han}, \bibinfo{person}{Bin Yin}, \bibinfo{person}{Shangyu Chen}, \bibinfo{person}{He Jiang}, \bibinfo{person}{Fei Jiang}, \bibinfo{person}{Xiang Li}, \bibinfo{person}{Chi Ma}, \bibinfo{person}{Mincong Huang}, \bibinfo{person}{Xiaoguang Li}, \bibinfo{person}{Chunzhen Jing}, \bibinfo{person}{Yueming Han}, \bibinfo{person}{Menglei Zhou}, \bibinfo{person}{Lei Yu}, \bibinfo{person}{Chuan Liu}, {and} \bibinfo{person}{Wei Lin}.} \bibinfo{year}{2025}\natexlab{}.
\newblock \bibinfo{title}{MTGR: Industrial-Scale Generative Recommendation Framework in Meituan}.
\newblock
\showeprint[arxiv]{2505.18654}~[cs.IR]
\urldef\tempurl%
\url{https://arxiv.org/abs/2505.18654}
\showURL{%
\tempurl}


\bibitem[He et~al\mbox{.}(2014)]%
        {he2014practical}
\bibfield{author}{\bibinfo{person}{Xinran He}, \bibinfo{person}{Junfeng Pan}, \bibinfo{person}{Ou Jin}, \bibinfo{person}{Tianbing Xu}, \bibinfo{person}{Bo Liu}, \bibinfo{person}{Tao Xu}, \bibinfo{person}{Yanxin Shi}, \bibinfo{person}{Antoine Atallah}, \bibinfo{person}{Ralf Herbrich}, \bibinfo{person}{Stuart Bowers}, {et~al\mbox{.}}} \bibinfo{year}{2014}\natexlab{}.
\newblock \showarticletitle{Practical lessons from predicting clicks on ads at facebook}. In \bibinfo{booktitle}{\emph{Proceedings of the eighth international workshop on data mining for online advertising}}. \bibinfo{pages}{1--9}.
\newblock


\bibitem[Huang et~al\mbox{.}(2020)]%
        {fbsearch_arxiv2020}
\bibfield{author}{\bibinfo{person}{Jui{-}Ting Huang}, \bibinfo{person}{Ashish Sharma}, \bibinfo{person}{Shuying Sun}, \bibinfo{person}{Li Xia}, \bibinfo{person}{David Zhang}, \bibinfo{person}{Philip Pronin}, \bibinfo{person}{Janani Padmanabhan}, \bibinfo{person}{Giuseppe Ottaviano}, {and} \bibinfo{person}{Linjun Yang}.} \bibinfo{year}{2020}\natexlab{}.
\newblock \showarticletitle{Embedding-based Retrieval in Facebook Search}.
\newblock \bibinfo{journal}{\emph{CoRR}}  \bibinfo{volume}{abs/2006.11632} (\bibinfo{year}{2020}).
\newblock
\showeprint[arXiv]{2006.11632}
\urldef\tempurl%
\url{https://arxiv.org/abs/2006.11632}
\showURL{%
\tempurl}


\bibitem[Huang et~al\mbox{.}(2025)]%
        {huangxhs25}
\bibfield{author}{\bibinfo{person}{Yanhua Huang}, \bibinfo{person}{Yuqi Chen}, \bibinfo{person}{Xiong Cao}, \bibinfo{person}{Rui Yang}, \bibinfo{person}{Mingliang Qi}, \bibinfo{person}{Yinghao Zhu}, \bibinfo{person}{Qingchang Han}, \bibinfo{person}{Yaowei Liu}, \bibinfo{person}{Zhaoyu Liu}, \bibinfo{person}{Xuefeng Yao}, \bibinfo{person}{Yuting Jia}, \bibinfo{person}{Leilei Ma}, \bibinfo{person}{Yinqi Zhang}, \bibinfo{person}{Taoyu Zhu}, \bibinfo{person}{Liujie Zhang}, \bibinfo{person}{Lei Chen}, \bibinfo{person}{Weihang Chen}, \bibinfo{person}{Min Zhu}, \bibinfo{person}{Ruiwen Xu}, {and} \bibinfo{person}{Lei Zhang}.} \bibinfo{year}{2025}\natexlab{}.
\newblock \bibinfo{title}{Towards Large-scale Generative Ranking}.
\newblock
\showeprint[arxiv]{2505.04180}~[cs.IR]
\urldef\tempurl%
\url{https://arxiv.org/abs/2505.04180}
\showURL{%
\tempurl}


\bibitem[Ivchenko et~al\mbox{.}(2022)]%
        {ivchenko2022torchrec}
\bibfield{author}{\bibinfo{person}{Dmytro Ivchenko}, \bibinfo{person}{Dennis Van Der~Staay}, \bibinfo{person}{Colin Taylor}, \bibinfo{person}{Xing Liu}, \bibinfo{person}{Will Feng}, \bibinfo{person}{Rahul Kindi}, \bibinfo{person}{Anirudh Sudarshan}, {and} \bibinfo{person}{Shahin Sefati}.} \bibinfo{year}{2022}\natexlab{}.
\newblock \showarticletitle{Torchrec: a pytorch domain library for recommendation systems}. In \bibinfo{booktitle}{\emph{Proceedings of the 16th ACM Conference on Recommender Systems}}. \bibinfo{pages}{482--483}.
\newblock


\bibitem[Jouppi et~al\mbox{.}(2017)]%
        {jouppi2017datacenter}
\bibfield{author}{\bibinfo{person}{Norman~P Jouppi}, \bibinfo{person}{Cliff Young}, \bibinfo{person}{Nishant Patil}, \bibinfo{person}{David Patterson}, \bibinfo{person}{Gaurav Agrawal}, \bibinfo{person}{Raminder Bajwa}, \bibinfo{person}{Sarah Bates}, \bibinfo{person}{Suresh Bhatia}, \bibinfo{person}{Nan Boden}, \bibinfo{person}{Al Borchers}, {et~al\mbox{.}}} \bibinfo{year}{2017}\natexlab{}.
\newblock \showarticletitle{In-datacenter performance analysis of a tensor processing unit}. In \bibinfo{booktitle}{\emph{Proceedings of the 44th annual international symposium on computer architecture}}. \bibinfo{pages}{1--12}.
\newblock


\bibitem[Kang and McAuley(2018)]%
        {kang2018self}
\bibfield{author}{\bibinfo{person}{Wang-Cheng Kang} {and} \bibinfo{person}{Julian McAuley}.} \bibinfo{year}{2018}\natexlab{}.
\newblock \showarticletitle{Self-attentive sequential recommendation}. In \bibinfo{booktitle}{\emph{2018 IEEE international conference on data mining (ICDM)}}. IEEE, \bibinfo{pages}{197--206}.
\newblock


\bibitem[Kent(1983)]%
        {kent1983simple}
\bibfield{author}{\bibinfo{person}{William Kent}.} \bibinfo{year}{1983}\natexlab{}.
\newblock \showarticletitle{A simple guide to five normal forms in relational database theory}.
\newblock \bibinfo{journal}{\emph{Commun. ACM}} \bibinfo{volume}{26}, \bibinfo{number}{2} (\bibinfo{year}{1983}), \bibinfo{pages}{120--125}.
\newblock


\bibitem[Khrylchenko et~al\mbox{.}(2025)]%
        {khrylchenko2025scalingrecommendertransformersbillion}
\bibfield{author}{\bibinfo{person}{Kirill Khrylchenko}, \bibinfo{person}{Artem Matveev}, \bibinfo{person}{Sergei Makeev}, {and} \bibinfo{person}{Vladimir Baikalov}.} \bibinfo{year}{2025}\natexlab{}.
\newblock \bibinfo{title}{Scaling Recommender Transformers to One Billion Parameters}.
\newblock
\showeprint[arxiv]{2507.15994}~[cs.IR]
\urldef\tempurl%
\url{https://arxiv.org/abs/2507.15994}
\showURL{%
\tempurl}


\bibitem[Li et~al\mbox{.}(2019)]%
        {mind_2019}
\bibfield{author}{\bibinfo{person}{Chao Li}, \bibinfo{person}{Zhiyuan Liu}, \bibinfo{person}{Mengmeng Wu}, \bibinfo{person}{Yuchi Xu}, \bibinfo{person}{Pipei Huang}, \bibinfo{person}{Huan Zhao}, \bibinfo{person}{Guoliang Kang}, \bibinfo{person}{Qiwei Chen}, \bibinfo{person}{Wei Li}, {and} \bibinfo{person}{Dik~Lun Lee}.} \bibinfo{year}{2019}\natexlab{}.
\newblock \bibinfo{title}{Multi-Interest Network with Dynamic Routing for Recommendation at Tmall}.
\newblock
\showeprint[arxiv]{1904.08030}~[cs.IR]
\urldef\tempurl%
\url{https://arxiv.org/abs/1904.08030}
\showURL{%
\tempurl}


\bibitem[Li et~al\mbox{.}(2022)]%
        {mlp4rec_ijcai22}
\bibfield{author}{\bibinfo{person}{Muyang Li}, \bibinfo{person}{Xiangyu Zhao}, \bibinfo{person}{Chuan Lyu}, \bibinfo{person}{Minghao Zhao}, \bibinfo{person}{Runze Wu}, {and} \bibinfo{person}{Ruocheng Guo}.} \bibinfo{year}{2022}\natexlab{}.
\newblock \showarticletitle{MLP4Rec: A Pure MLP Architecture for Sequential Recommendations}. In \bibinfo{booktitle}{\emph{Proceedings of the Thirty-First International Joint Conference on Artificial Intelligence, {IJCAI-22}}}, \bibfield{editor}{\bibinfo{person}{Lud~De Raedt}} (Ed.). \bibinfo{publisher}{International Joint Conferences on Artificial Intelligence Organization}, \bibinfo{pages}{2138--2144}.
\newblock
\href{https://doi.org/10.24963/ijcai.2022/297}{doi:\nolinkurl{10.24963/ijcai.2022/297}}
\newblock
\shownote{Main Track}.


\bibitem[Liu et~al\mbox{.}(2024)]%
        {liu2024kuaiformer}
\bibfield{author}{\bibinfo{person}{Chi Liu}, \bibinfo{person}{Jiangxia Cao}, \bibinfo{person}{Rui Huang}, \bibinfo{person}{Kai Zheng}, \bibinfo{person}{Qiang Luo}, \bibinfo{person}{Kun Gai}, {and} \bibinfo{person}{Guorui Zhou}.} \bibinfo{year}{2024}\natexlab{}.
\newblock \showarticletitle{KuaiFormer: Transformer-Based Retrieval at Kuaishou}.
\newblock \bibinfo{journal}{\emph{arXiv preprint arXiv:2411.10057}} (\bibinfo{year}{2024}).
\newblock


\bibitem[Lv et~al\mbox{.}(2025)]%
        {lv2025marm}
\bibfield{author}{\bibinfo{person}{Xiao Lv}, \bibinfo{person}{Jiangxia Cao}, \bibinfo{person}{Shijie Guan}, \bibinfo{person}{Xiaoyou Zhou}, \bibinfo{person}{Zhiguang Qi}, \bibinfo{person}{Yaqiang Zang}, \bibinfo{person}{Ming Li}, \bibinfo{person}{Ben Wang}, \bibinfo{person}{Kun Gai}, {and} \bibinfo{person}{Guorui Zhou}.} \bibinfo{year}{2025}\natexlab{}.
\newblock \bibinfo{title}{MARM: Unlocking the Future of Recommendation Systems through Memory Augmentation and Scalable Complexity}.
\newblock
\showeprint[arxiv]{2411.09425}~[cs.IR]
\urldef\tempurl%
\url{https://arxiv.org/abs/2411.09425}
\showURL{%
\tempurl}


\bibitem[Mudigere et~al\mbox{.}(2022)]%
        {mudigere2022software}
\bibfield{author}{\bibinfo{person}{Dheevatsa Mudigere}, \bibinfo{person}{Yuchen Hao}, \bibinfo{person}{Jianyu Huang}, \bibinfo{person}{Zhihao Jia}, \bibinfo{person}{Andrew Tulloch}, \bibinfo{person}{Srinivas Sridharan}, \bibinfo{person}{Xing Liu}, \bibinfo{person}{Mustafa Ozdal}, \bibinfo{person}{Jade Nie}, \bibinfo{person}{Jongsoo Park}, {et~al\mbox{.}}} \bibinfo{year}{2022}\natexlab{}.
\newblock \showarticletitle{Software-hardware co-design for fast and scalable training of deep learning recommendation models}. In \bibinfo{booktitle}{\emph{Proceedings of the 49th Annual International Symposium on Computer Architecture}}. \bibinfo{pages}{993--1011}.
\newblock


\bibitem[Naumov et~al\mbox{.}(2019)]%
        {naumov2019deep}
\bibfield{author}{\bibinfo{person}{Maxim Naumov}, \bibinfo{person}{Dheevatsa Mudigere}, \bibinfo{person}{Hao-Jun~Michael Shi}, \bibinfo{person}{Jianyu Huang}, \bibinfo{person}{Narayanan Sundaraman}, \bibinfo{person}{Jongsoo Park}, \bibinfo{person}{Xiaodong Wang}, \bibinfo{person}{Udit Gupta}, \bibinfo{person}{Carole-Jean Wu}, \bibinfo{person}{Alisson~G Azzolini}, {et~al\mbox{.}}} \bibinfo{year}{2019}\natexlab{}.
\newblock \showarticletitle{Deep learning recommendation model for personalization and recommendation systems}.
\newblock \bibinfo{journal}{\emph{arXiv preprint arXiv:1906.00091}} (\bibinfo{year}{2019}).
\newblock


\bibitem[Pancha et~al\mbox{.}(2022)]%
        {pancha2022pinnerformer}
\bibfield{author}{\bibinfo{person}{Nikil Pancha}, \bibinfo{person}{Andrew Zhai}, \bibinfo{person}{Jure Leskovec}, {and} \bibinfo{person}{Charles Rosenberg}.} \bibinfo{year}{2022}\natexlab{}.
\newblock \showarticletitle{Pinnerformer: Sequence modeling for user representation at pinterest}. In \bibinfo{booktitle}{\emph{Proceedings of the 28th ACM SIGKDD conference on knowledge discovery and data mining}}. \bibinfo{pages}{3702--3712}.
\newblock


\bibitem[Pi et~al\mbox{.}(2019)]%
        {pi2019practice}
\bibfield{author}{\bibinfo{person}{Qi Pi}, \bibinfo{person}{Weijie Bian}, \bibinfo{person}{Guorui Zhou}, \bibinfo{person}{Xiaoqiang Zhu}, {and} \bibinfo{person}{Kun Gai}.} \bibinfo{year}{2019}\natexlab{}.
\newblock \showarticletitle{Practice on long sequential user behavior modeling for click-through rate prediction}. In \bibinfo{booktitle}{\emph{Proceedings of the 25th ACM SIGKDD International Conference on Knowledge Discovery \& Data Mining}}. \bibinfo{pages}{2671--2679}.
\newblock


\bibitem[Pi et~al\mbox{.}(2020)]%
        {pi2020search}
\bibfield{author}{\bibinfo{person}{Qi Pi}, \bibinfo{person}{Guorui Zhou}, \bibinfo{person}{Yujing Zhang}, \bibinfo{person}{Zhe Wang}, \bibinfo{person}{Lejian Ren}, \bibinfo{person}{Ying Fan}, \bibinfo{person}{Xiaoqiang Zhu}, {and} \bibinfo{person}{Kun Gai}.} \bibinfo{year}{2020}\natexlab{}.
\newblock \showarticletitle{Search-based user interest modeling with lifelong sequential behavior data for click-through rate prediction}. In \bibinfo{booktitle}{\emph{Proceedings of the 29th ACM International Conference on Information \& Knowledge Management}}. \bibinfo{pages}{2685--2692}.
\newblock


\bibitem[Rajput et~al\mbox{.}(2023)]%
        {tiger_rajput2023recommender}
\bibfield{author}{\bibinfo{person}{Shashank Rajput}, \bibinfo{person}{Nikhil Mehta}, \bibinfo{person}{Anima Singh}, \bibinfo{person}{Raghunandan~Hulikal Keshavan}, \bibinfo{person}{Trung Vu}, \bibinfo{person}{Lukasz Heldt}, \bibinfo{person}{Lichan Hong}, \bibinfo{person}{Yi Tay}, \bibinfo{person}{Vinh~Q. Tran}, \bibinfo{person}{Jonah Samost}, \bibinfo{person}{Maciej Kula}, \bibinfo{person}{Ed~H. Chi}, {and} \bibinfo{person}{Maheswaran Sathiamoorthy}.} \bibinfo{year}{2023}\natexlab{}.
\newblock \showarticletitle{Recommender Systems with Generative Retrieval}. In \bibinfo{booktitle}{\emph{Thirty-seventh Conference on Neural Information Processing Systems}}.
\newblock
\urldef\tempurl%
\url{https://openreview.net/forum?id=BJ0fQUU32w}
\showURL{%
\tempurl}


\bibitem[Si et~al\mbox{.}(2024)]%
        {si2024twin}
\bibfield{author}{\bibinfo{person}{Zihua Si}, \bibinfo{person}{Lin Guan}, \bibinfo{person}{ZhongXiang Sun}, \bibinfo{person}{Xiaoxue Zang}, \bibinfo{person}{Jing Lu}, \bibinfo{person}{Yiqun Hui}, \bibinfo{person}{Xingchao Cao}, \bibinfo{person}{Zeyu Yang}, \bibinfo{person}{Yichen Zheng}, \bibinfo{person}{Dewei Leng}, {et~al\mbox{.}}} \bibinfo{year}{2024}\natexlab{}.
\newblock \showarticletitle{Twin v2: Scaling ultra-long user behavior sequence modeling for enhanced ctr prediction at kuaishou}. In \bibinfo{booktitle}{\emph{Proceedings of the 33rd ACM International Conference on Information and Knowledge Management}}. \bibinfo{pages}{4890--4897}.
\newblock


\bibitem[Tolstikhin et~al\mbox{.}(2021)]%
        {mlpmixer_neurips21}
\bibfield{author}{\bibinfo{person}{Ilya Tolstikhin}, \bibinfo{person}{Neil Houlsby}, \bibinfo{person}{Alexander Kolesnikov}, \bibinfo{person}{Lucas Beyer}, \bibinfo{person}{Xiaohua Zhai}, \bibinfo{person}{Thomas Unterthiner}, \bibinfo{person}{Jessica Yung}, \bibinfo{person}{Andreas Steiner}, \bibinfo{person}{Daniel Keysers}, \bibinfo{person}{Jakob Uszkoreit}, \bibinfo{person}{Mario Lucic}, {and} \bibinfo{person}{Alexey Dosovitskiy}.} \bibinfo{year}{2021}\natexlab{}.
\newblock \showarticletitle{MLP-mixer: an all-MLP architecture for vision}. In \bibinfo{booktitle}{\emph{Proceedings of the 35th International Conference on Neural Information Processing Systems}} \emph{(\bibinfo{series}{NIPS '21})}. \bibinfo{publisher}{Curran Associates Inc.}, \bibinfo{address}{Red Hook, NY, USA}, Article \bibinfo{articleno}{1857}, \bibinfo{numpages}{12}~pages.
\newblock
\showISBNx{9781713845393}


\bibitem[Vaswani(2017)]%
        {vaswani2017attention}
\bibfield{author}{\bibinfo{person}{A Vaswani}.} \bibinfo{year}{2017}\natexlab{}.
\newblock \showarticletitle{Attention is all you need}.
\newblock \bibinfo{journal}{\emph{Advances in Neural Information Processing Systems}} (\bibinfo{year}{2017}).
\newblock


\bibitem[Wang et~al\mbox{.}(2021)]%
        {wang2021dcn}
\bibfield{author}{\bibinfo{person}{Ruoxi Wang}, \bibinfo{person}{Rakesh Shivanna}, \bibinfo{person}{Derek Cheng}, \bibinfo{person}{Sagar Jain}, \bibinfo{person}{Dong Lin}, \bibinfo{person}{Lichan Hong}, {and} \bibinfo{person}{Ed Chi}.} \bibinfo{year}{2021}\natexlab{}.
\newblock \showarticletitle{Dcn v2: Improved deep \& cross network and practical lessons for web-scale learning to rank systems}. In \bibinfo{booktitle}{\emph{Proceedings of the web conference 2021}}. \bibinfo{pages}{1785--1797}.
\newblock


\bibitem[Wilkening et~al\mbox{.}(2021)]%
        {wilkening2021recssd}
\bibfield{author}{\bibinfo{person}{Mark Wilkening}, \bibinfo{person}{Udit Gupta}, \bibinfo{person}{Samuel Hsia}, \bibinfo{person}{Caroline Trippel}, \bibinfo{person}{Carole-Jean Wu}, \bibinfo{person}{David Brooks}, {and} \bibinfo{person}{Gu-Yeon Wei}.} \bibinfo{year}{2021}\natexlab{}.
\newblock \showarticletitle{RecSSD: near data processing for solid state drive based recommendation inference}. In \bibinfo{booktitle}{\emph{Proceedings of the 26th ACM International Conference on Architectural Support for Programming Languages and Operating Systems}}. \bibinfo{pages}{717--729}.
\newblock


\bibitem[Xu et~al\mbox{.}(2022)]%
        {xu2022rethinking}
\bibfield{author}{\bibinfo{person}{Jiajing Xu}, \bibinfo{person}{Andrew Zhai}, {and} \bibinfo{person}{Charles Rosenberg}.} \bibinfo{year}{2022}\natexlab{}.
\newblock \showarticletitle{Rethinking personalized ranking at Pinterest: An end-to-end approach}. In \bibinfo{booktitle}{\emph{Proceedings of the 16th ACM Conference on Recommender Systems}}. \bibinfo{pages}{502--505}.
\newblock


\bibitem[You et~al\mbox{.}(2024)]%
        {you2024exploiting}
\bibfield{author}{\bibinfo{person}{Xin You}, \bibinfo{person}{Hailong Yang}, \bibinfo{person}{Siqi Wang}, \bibinfo{person}{Tao Peng}, \bibinfo{person}{Chen Ding}, \bibinfo{person}{Xinyuan Li}, \bibinfo{person}{Bangduo Chen}, \bibinfo{person}{Zhongzhi Luan}, \bibinfo{person}{Tongxuan Liu}, \bibinfo{person}{Yong Li}, {and} \bibinfo{person}{Depei Qian}.} \bibinfo{year}{2024}\natexlab{}.
\newblock \showarticletitle{Exploiting Structured Feature and Runtime Isolation for High-Performant Recommendation Serving}.
\newblock \bibinfo{journal}{\emph{IEEE Trans. Comput.}} \bibinfo{volume}{73}, \bibinfo{number}{11} (\bibinfo{year}{2024}), \bibinfo{pages}{2474--2487}.
\newblock
\href{https://doi.org/10.1109/TC.2024.3449749}{doi:\nolinkurl{10.1109/TC.2024.3449749}}


\bibitem[Zhai et~al\mbox{.}(2023)]%
        {zhai2023revisiting}
\bibfield{author}{\bibinfo{person}{Jiaqi Zhai}, \bibinfo{person}{Zhaojie Gong}, \bibinfo{person}{Yueming Wang}, \bibinfo{person}{Xiao Sun}, \bibinfo{person}{Zheng Yan}, \bibinfo{person}{Fu Li}, {and} \bibinfo{person}{Xing Liu}.} \bibinfo{year}{2023}\natexlab{}.
\newblock \showarticletitle{Revisiting Neural Retrieval on Accelerators}. In \bibinfo{booktitle}{\emph{Proceedings of the 29th ACM SIGKDD Conference on Knowledge Discovery and Data Mining}}. \bibinfo{pages}{5520--5531}.
\newblock


\bibitem[Zhai et~al\mbox{.}(2024)]%
        {zhai2024actions}
\bibfield{author}{\bibinfo{person}{Jiaqi Zhai}, \bibinfo{person}{Lucy Liao}, \bibinfo{person}{Xing Liu}, \bibinfo{person}{Yueming Wang}, \bibinfo{person}{Rui Li}, \bibinfo{person}{Xuan Cao}, \bibinfo{person}{Leon Gao}, \bibinfo{person}{Zhaojie Gong}, \bibinfo{person}{Fangda Gu}, \bibinfo{person}{Michael He}, {et~al\mbox{.}}} \bibinfo{year}{2024}\natexlab{}.
\newblock \showarticletitle{Actions speak louder than words: Trillion-parameter sequential transducers for generative recommendations}.
\newblock \bibinfo{journal}{\emph{arXiv preprint arXiv:2402.17152}} (\bibinfo{year}{2024}).
\newblock


\bibitem[Zhang et~al\mbox{.}(2022)]%
        {zhang2022dhen}
\bibfield{author}{\bibinfo{person}{Buyun Zhang}, \bibinfo{person}{Liang Luo}, \bibinfo{person}{Xi Liu}, \bibinfo{person}{Jay Li}, \bibinfo{person}{Zeliang Chen}, \bibinfo{person}{Weilin Zhang}, \bibinfo{person}{Xiaohan Wei}, \bibinfo{person}{Yuchen Hao}, \bibinfo{person}{Michael Tsang}, \bibinfo{person}{Wenjun Wang}, {et~al\mbox{.}}} \bibinfo{year}{2022}\natexlab{}.
\newblock \showarticletitle{DHEN: A deep and hierarchical ensemble network for large-scale click-through rate prediction}.
\newblock \bibinfo{journal}{\emph{arXiv preprint arXiv:2203.11014}} (\bibinfo{year}{2022}).
\newblock


\bibitem[Zhang et~al\mbox{.}(2025)]%
        {killingbirdsstoneunifying_sigir25}
\bibfield{author}{\bibinfo{person}{Luankang Zhang}, \bibinfo{person}{Kenan Song}, \bibinfo{person}{Yi~Quan Lee}, \bibinfo{person}{Wei Guo}, \bibinfo{person}{Hao Wang}, \bibinfo{person}{Yawen Li}, \bibinfo{person}{Huifeng Guo}, \bibinfo{person}{Yong Liu}, \bibinfo{person}{Defu Lian}, {and} \bibinfo{person}{Enhong Chen}.} \bibinfo{year}{2025}\natexlab{}.
\newblock \bibinfo{title}{Killing Two Birds with One Stone: Unifying Retrieval and Ranking with a Single Generative Recommendation Model}.
\newblock
\showeprint[arxiv]{2504.16454}~[cs.IR]
\urldef\tempurl%
\url{https://arxiv.org/abs/2504.16454}
\showURL{%
\tempurl}


\bibitem[Zhao et~al\mbox{.}(2022)]%
        {zhao2022understanding}
\bibfield{author}{\bibinfo{person}{Mark Zhao}, \bibinfo{person}{Niket Agarwal}, \bibinfo{person}{Aarti Basant}, \bibinfo{person}{Bu{\u{g}}ra Gedik}, \bibinfo{person}{Satadru Pan}, \bibinfo{person}{Mustafa Ozdal}, \bibinfo{person}{Rakesh Komuravelli}, \bibinfo{person}{Jerry Pan}, \bibinfo{person}{Tianshu Bao}, \bibinfo{person}{Haowei Lu}, {et~al\mbox{.}}} \bibinfo{year}{2022}\natexlab{}.
\newblock \showarticletitle{Understanding data storage and ingestion for large-scale deep recommendation model training: Industrial product}. In \bibinfo{booktitle}{\emph{Proceedings of the 49th annual international symposium on computer architecture}}. \bibinfo{pages}{1042--1057}.
\newblock


\bibitem[Zhao et~al\mbox{.}(2023)]%
        {zhao2023recd}
\bibfield{author}{\bibinfo{person}{Mark Zhao}, \bibinfo{person}{Dhruv Choudhary}, \bibinfo{person}{Devashish Tyagi}, \bibinfo{person}{Ajay Somani}, \bibinfo{person}{Max Kaplan}, \bibinfo{person}{Sung-Han Lin}, \bibinfo{person}{Sarunya Pumma}, \bibinfo{person}{Jongsoo Park}, \bibinfo{person}{Aarti Basant}, \bibinfo{person}{Niket Agarwal}, {et~al\mbox{.}}} \bibinfo{year}{2023}\natexlab{}.
\newblock \showarticletitle{RecD: Deduplication for end-to-end deep learning recommendation model training infrastructure}.
\newblock \bibinfo{journal}{\emph{Proceedings of Machine Learning and Systems}}  \bibinfo{volume}{5} (\bibinfo{year}{2023}), \bibinfo{pages}{754--767}.
\newblock


\bibitem[Zhou et~al\mbox{.}(2019)]%
        {zhou2019deep}
\bibfield{author}{\bibinfo{person}{Guorui Zhou}, \bibinfo{person}{Na Mou}, \bibinfo{person}{Ying Fan}, \bibinfo{person}{Qi Pi}, \bibinfo{person}{Weijie Bian}, \bibinfo{person}{Chang Zhou}, \bibinfo{person}{Xiaoqiang Zhu}, {and} \bibinfo{person}{Kun Gai}.} \bibinfo{year}{2019}\natexlab{}.
\newblock \showarticletitle{Deep interest evolution network for click-through rate prediction}. In \bibinfo{booktitle}{\emph{Proceedings of the AAAI conference on artificial intelligence}}, Vol.~\bibinfo{volume}{33}. \bibinfo{pages}{5941--5948}.
\newblock


\bibitem[Zhou et~al\mbox{.}(2022)]%
        {zhou2022serving}
\bibfield{author}{\bibinfo{person}{Lixi Zhou}, \bibinfo{person}{Jiaqing Chen}, \bibinfo{person}{Amitabh Das}, \bibinfo{person}{Hong Min}, \bibinfo{person}{Lei Yu}, \bibinfo{person}{Ming Zhao}, {and} \bibinfo{person}{Jia Zou}.} \bibinfo{year}{2022}\natexlab{}.
\newblock \showarticletitle{Serving deep learning models with deduplication from relational databases}.
\newblock \bibinfo{journal}{\emph{arXiv preprint arXiv:2201.10442}} (\bibinfo{year}{2022}).
\newblock


\bibitem[Zuo et~al\mbox{.}(2024)]%
        {ccs_kdd24}
\bibfield{author}{\bibinfo{person}{Chandler Zuo}, \bibinfo{person}{Jonathan Castaldo}, \bibinfo{person}{Hanqing Zhu}, \bibinfo{person}{Haoyu Zhang}, \bibinfo{person}{Ji Liu}, \bibinfo{person}{Yangpeng Ou}, {and} \bibinfo{person}{Xiao Kong}.} \bibinfo{year}{2024}\natexlab{}.
\newblock \showarticletitle{Inductive Modeling for Realtime Cold Start Recommendations}. In \bibinfo{booktitle}{\emph{Proceedings of the 30th ACM SIGKDD Conference on Knowledge Discovery and Data Mining}} (Barcelona, Spain) \emph{(\bibinfo{series}{KDD '24})}. \bibinfo{publisher}{Association for Computing Machinery}, \bibinfo{address}{New York, NY, USA}, \bibinfo{pages}{6400–6409}.
\newblock
\showISBNx{9798400704901}
\href{https://doi.org/10.1145/3637528.3671588}{doi:\nolinkurl{10.1145/3637528.3671588}}


\end{thebibliography}

\newpage

\appendix
\section{Notations}
We summarize key notations used in this paper in Table~\ref{tbl:table-of-notations}.

\begin{table*}[h]
\caption{Table of Notations.}
\vspace{-.5em}
\label{tbl:table-of-notations}
\begin{center}
\begin{tabular}[p]{cl}
\toprule
\bf Symbol & \bf Description \\
\midrule
$num\_candidate$& number of candidates in a request \\
\midrule
$B_{NRO}$ & Number of impression-level samples (i.e., number of items) in a training batch. \\
$B_{RO}$ & Number of request-level samples (i.e., number of requests) in a training batch.\\
\midrule
$RO_i$ & $i$-th RO-side impression in the user history list.\\
$NRO_i$ & $i$-th target item in the request.\\
$a$ & Action feature.\\
$c$ & Context features, for instance surface type, timestamp, etc.\\
$MT$ & Multi-task prediction for ranking models.\\
$u$ & User representation aggregated from user history sequence. \\
$\Phi$ & Item served in product. \\
\midrule
$X$ & Input tensor of sparse feature embeddings.\\
$B$ & Batch size. Note that this is invariant to ROO.\\
$d_{in}$ & Input dimension of sparse feature embeddings. \\
$n_{in}$ & Input number of sparse feature embeddings. \\
$d_{out}$ & Output dimension of sparse feature embeddings. \\
$n_{out}$ & Output number of sparse feature embeddings. \\
\bottomrule
\end{tabular}
\end{center}
\end{table*}

\section{Request Level Join}
\label{app:request-level-join}
We provide pseudocode to explain request-level joiner and ROO training data. Compared to impression-level joiners such as RecD \cite{zhao2023recd}, the design and implementation details set ROO apart.
\begin{itemize}
    \item The request level joiner supports different triggers to close a join window and publish a ROO training sample. In production, the average close time of the join window is 16 minutes and satisfies the model activity-to-serving requirement.
    \item The request level joiner keeps a single instance of user-side features, e.g., roIdListFeatures (line 7), for all impressions of the same request in the in-memory feature storage of the joiner. Consequently, it actually uses much less memory than impression-level joiners.
    \item Due to feature deduplication at the root data source of the training data pipeline, the ROO training data does not require additional data infra feature, nor does it add any data landing latency increase. In contrast, RecD requires extra ETL jobs to cluster samples by session, which adds data landing latency by 30 minutes in production and makes it inherently incompatible for online training and model responsiveness. ROO data has a notable advantage over RecD.
    \item The distinction of RO and NRO features in ROO data is leveraged by feature preprocessing to produce 2D tensor representations as PyTorch KeyedJaggedTensor. In contrast, RecD uses a custom InverseKeyedJaggeredTensor to represent impression samples of the same request. The custom tensor transformation requires 1.5x more data preprocessing cpu overhead than ROO for the same GPU unit.
\end{itemize}

\begin{algorithm}[h] \label{algo:roo}
  \caption{Request Level Join}
  \begin{algorithmic}[1]
  \State Struct RequestJoinRecord \{
    \State Int userId;
    \State Int requestId;
    \State Set<Int> impressions;\Comment{impression items}
    \State Map<Int, List<Int>> conversions;\Comment{item labels}
    \State Map<Int, Float> roFloatFeatures;
    \State Map<Int, List<Int>> roIdListFeatures;
    \State Map<Int, Map<Int, Float>> nroFloatFeatures;
    \State Map<Int, Map<Int, List<Int>>> nroIdListFeatures;
  \State \}
  \State
  \Procedure {RequestLevelJoin}{$userId, requestId, payload$}:
    \State currRequestId = getCurrentJoinerRequestId(userId)  
    \State joinKey = JoinKey(userId, currRequestId)
    \State
    \If {$userHasNewRequestId(requestId, currRequestId)$ || $exceedsUserEngagementThreashold(joinKey)$ || $joinWindowTimeUp(joinKey)$}
      \State closeJoinWindow(joinKey)\Comment{publish ROO sample}
      \State shouldStartNewJoinWindow = true
    \EndIf
    \State
    \If {$shouldStartNewJoinWindow$}
      \State rec = RequestJoinRecord()\Comment{new join record}
      \State rec.impressions.add($payload.item$)
      \State setRoFeatures($rec, userId, requestId$)
      \State Joiner.set(JoinKey($userId, requestId$), $rec$)
    \Else{}
      \State rec = Joiner.get($joinKey$)\Comment{update join record}
      \If {$rec.impressions.contains(payload.item)$}
        \State rec.conversions[$payload.item$].add($payload.itemLabels$)
      \Else{}
        \State rec.impressions.add($payload.item$)
        \State setNroFeatures($rec, payload.item, requestId$)
      \EndIf
      \State Joiner.set($joinKey, rec$)
    \EndIf
  \EndProcedure
  \end{algorithmic}
\end{algorithm}

\section{ROO Expansion for Backward Compatibility}
Despite its significant storage efficiency advantage, request-level training sample schema is incompatible with the traditional DLRMs trained with impression-level schema. In a typical industrial environment, hundreds of models across multiple ranking stages collectively serve billions of users using a product, and not all models have to adopt the ROO training paradigm due to return-on-investment (ROI) considerations. We present a practical way to minimize the migration cost for these models and facilitate the adoption of ROO data schema. We introduce ROO expansion data adapter in the feature preprocessing layer. 

For each mini-batch, the data adapter expands a ROO sample into multiple in-memory impression samples, with user-side features duplicated to the in-memory data frames. Trainer side rebatching is leveraged to keep the consistent training batch size. This equips ROO data schema with backward compatibility and eliminates unnecessary model rewrite efforts. 
Note that the ROO expansion requires more memory and computation cost on the data preprocessing workers. However, reading ROO samples has a 10x drop IO bytes from the warm storage (Hive) as compared to reading impression samples. Despite the expansion overhead, it had a 2-3x data preprocessing throughput increase.
The expansion functionality simplifies the migration process; major products fully migrated all retrieval, early stage ranking and late stage ranking models to ROO data within a 9-month time frame.

\section{Ranking Model Architecture}
\begin{figure*}[h]
  \centering
  \includegraphics[width=0.7\linewidth]{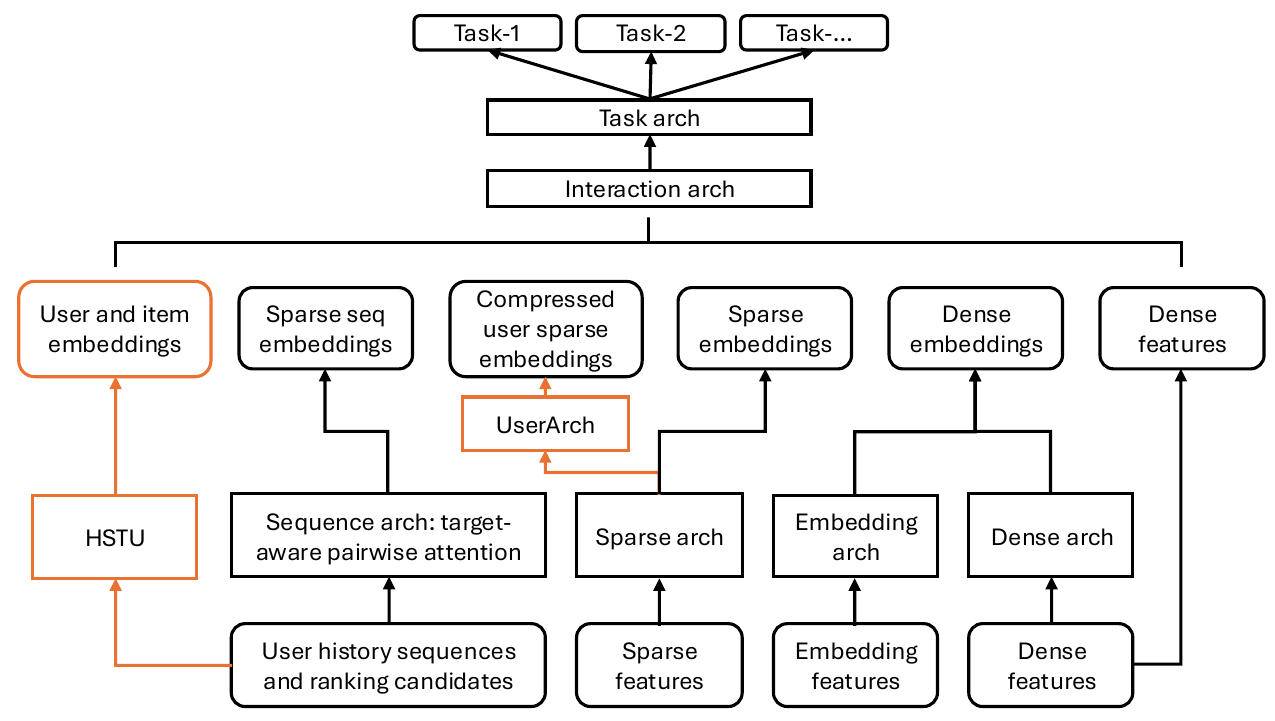}
  \caption{The late stage ranking model architecture. HSTU and user arch (highlighted) introduced request only computations to traditional DLRM architecture.}
  \Description{The late stage ranking model architecture. HSTU and user arch (highlighted) introduced additional RO computations.}
  \label{fig:lsr_model_arch}
\end{figure*}

We illustrate a common DLRM-based ranking model baseline architecture discussed in this paper in Figure~\ref{fig:lsr_model_arch}.

\newpage

\end{document}